\documentclass[12pt,a4paper,twoside,fleqn]{article}
\usepackage{graphicx,cite,feynmp}

\def\be{\begin{equation}}
\def\ee{\end{equation}}
\def\bea{\begin{eqnarray}}
\def\eea{\end{eqnarray}}
\def\nnb{\nonumber}
\def\bbuildrel#1_#2^#3{\mathrel{\mathop{\kern 0pt#1}\limits_{#2}^{#3}}}
\def\slash#1{\setbox0=\hbox{$#1$}#1\hskip-\wd0\dimen0=5pt\advance
       \dimen0 by-\ht0\advance\dimen0 by\dp0\lower0.5\dimen0\hbox
         to\wd0{\hss\sl/\/\hss}}
\def\gev{{\rm GeV}}
\def\mev{{\rm MeV}}

\newcommand{\gae}{\lower 2pt \hbox{$\, \buildrel {\scriptstyle >}\over
    {\scriptstyle \sim}\,$}}
\newcommand{\lae}{\lower 2pt \hbox{$\, \buildrel {\scriptstyle <}\over
    {\scriptstyle \sim}\,$}}

\newcommand{\scs}{\scriptscriptstyle}
\newcommand{\f}{\frac}
\newcommand{\fm}[2]{{\textstyle \frac{#1}{#2}}}
\newcommand{\me}[1]{\langle#1\rangle}

\newcommand{\aem}{\widetilde{\alpha}_{\mathrm e}}
\newcommand{\as}{\widetilde{\alpha}_{\mathrm s}}
\newcommand{\sot}{\hat{s}_{12}}
\newcommand{\s}{\hat{s}}
\newcommand{\se}{\scriptstyle}
\newcommand{\spp}{\vphantom{\bigg(}}

\begin{document}

\begin{titlepage}

\begin{flushright}
ZU-TH 06/05\\
IFT-6/2005\\
FERMILAB-PUB-05-531-T\\
hep-ph/0512066\\[2cm]
\end{flushright}

\begin{center}
\setlength {\baselineskip}{0.3in} {\bf\Large 
Electromagnetic Logarithms in $\bar{B} \to X_s \ell^+ \ell^-$}\\[2cm]

\setlength {\baselineskip}{0.2in} {\large Tobias Huber$^1$, Enrico
Lunghi$^{1,2}$, Miko{\l}aj Misiak$^{1,3}$ and Daniel Wyler$^1$}\\[5mm]

$^1$~{\it Institute for Theoretical Physics, University of Zurich,\\
          Winterthurerstrasse 190, CH-8057, Zurich, Switzerland.}\\[5mm]
          
$^2$~{\it Fermi National Accelerator Laboratory,\\
         P.O. Box 500, Batavia, IL 60510, U.S.A.}\\[5mm]

$^3$~{\it Institute of Theoretical Physics, Warsaw University,\\
          Ho\.za 69, PL-00-681 Warsaw, Poland.}\\[2cm] 

{\bf Abstract}\\[5mm]
\end{center} 
\setlength{\baselineskip}{0.2in} 

The $\bar{B} \to X_s \ell^+ \ell^-$ decay rate is known at the
next-to-next-to-leading order in QCD. It is proportional to $\alpha_{\mathrm
  em}(\mu)^2$ and has a $\pm 4$\% scale uncertainty before including the
${\cal O}\left(\alpha_{\mathrm em} \ln(M_W^2/m_b^2)\right)$ electromagnetic
corrections. We evaluate these corrections and confirm the earlier findings of
Bobeth {\it et al.}.  Furthermore, we complete the calculation of
logarithmically enhanced electromagnetic effects by including also
 QED corrections to the matrix elements of four-fermion
operators. Such corrections contain a  collinear logarithm
$\ln(m_b^2/m_\ell^2)$ that survives integration over the low dilepton
invariant mass region $1\;\gev^2 < m^2_{\ell\ell} < 6\;\gev^2$  and
  enhances the integrated decay rate in this domain. For the 
low-$m^2_{\ell\ell}$ integrated branching ratio  in the muonic case, we
  find  $ {\cal B} (B\to X_s \mu^+\mu^-) =   (1.59\pm 0.11)\times 10^{-6} $,
 where the error includes the parametric and perturbative uncertainties only.
  For ${\cal B} (B\to X_s e^+e^-)$, in the current BaBar and Belle setups,
  the logarithm of the lepton mass gets replaced by angular cut
  parameters and the integrated branching ratio for the electrons is
  expected to be close to that for the muons. 

\end{titlepage}

\section{Introduction}

The inclusive decay $\bar{B} \to X_s \ell^+ \ell^-$ with $l =
e$~or~$\mu$ is known to be a sensitive probe of new physics at the
electroweak scale.  Its branching ratio has been recently measured by
both Belle~\cite{Iwasaki:2005sy} and BaBar~\cite{Aubert:2004it}. In the
low dilepton invariant mass region, $1\;\gev^2 < m^2_{\ell\ell} <
6\;\gev^2$, the experimental results read
\bea
{\cal B} (B\to X_s \ell^+\ell^-) &=& (1.493 \pm 0.504^{+0.411}_{-0.321})
\times 10^{-6} \;\;\; ({\rm Belle}) \; ,\\ 
{\cal B} (B\to X_s \ell^+\ell^-) &=& (1.8 \pm 0.7\pm0.5)
\times 10^{-6} \;\;\; ({\rm BaBar}) \; .
\eea
This leads to a world average
\bea
{\cal B} (B\to X_s \ell^+\ell^-) &=& (1.60 \pm 0.51)\times 10^{-6} \; .
\eea
Measurements for lower and higher values of $m^2_{\ell\ell}$ are
available, too. However, for higher $m^2_{\ell\ell}$, non-perturbative
effects of the $J/\Psi$, $\Psi'$ and higher resonances are sizeable,
and the theoretical predictions have larger uncertainties. On the
other hand, for $m^2_{\ell\ell} < 1\;\gev^2$, the branching ratio is
determined largely by the contribution from almost real intermediate
photons, and it contains essentially the same information on new
physics as is already known from the $\bar{B} \to X_s \gamma$
measurements. Throughout this paper, we restrict ourselves to the
dilepton mass region $m_{\ell^+ \ell^-}^2 \in [1, 6]\;\gev^2$.

The experimental errors in the branching ratio are expected to be
substantially reduced in the near future. On the theoretical side, the
predictions are quite well under control because the inclusive hadronic
$\bar{B} \to X_s \ell^+ \ell^-$ decay rate for low dilepton mass is well
approximated by the perturbatively calculable partonic $b \to X_s^{\rm parton}
\ell^+ \ell^-$ decay rate. Thanks to the recent (practically) complete
calculation~\cite{Bobeth:1999mk, Asatryan:2001zw, Asatryan:2002iy,
  Ghinculov:2003qd, Gambino:2003zm, Gorbahn:2004my,Bobeth:2003at} of the
Next-to-Next-to-Leading Order (NNLO) QCD corrections, the perturbative
uncertainties are now below 10\%.

The branching ratio is proportional to $\alpha_{\mathrm em}^2(\mu)$
whose scale dependence cannot be neglected.  Indeed, at the leading
order in QED, ${\cal B} (B\to X_s \ell^+\ell^-)$ changes from
 $1.54\cdot10^{-6}$ to $1.65\cdot10^{-6}$  when the renormalization
scale of $\alpha_{\rm em}$ is changed from $\mu = {\cal O}(m_b)$ to
$\mu = {\cal O}(M_W)$.  This uncertainty is removed by calculating
those QED corrections that are enhanced by large logarithms
$\ln(M_H^2/M_L^2)$, where $M_H \sim M_W, m_t$ and $M_L \sim m_b,
m_{\ell\ell}$.

In Ref.~\cite{Bobeth:2003at}, the QED corrections to the Wilson
coefficients were calculated, thereby giving most of the
electromagnetic corrections that are enhanced by
$\ln(M_H^2/M_L^2)$. As a result, the authors find a branching ratio of
 $1.56\cdot 10^{-6},$ \footnote{
 The number quoted by the authors of Ref.~\cite{Bobeth:2003at} and on
which we agree is $1.57\cdot 10^{-6}$.  In the text we give the
result obtained using the updated experimental inputs summarized in
Table~\ref{tab:inputs}. }
which incidentally corresponds to setting $\alpha_{\rm em}^2 = \alpha_{\rm
em}^2(\mu \sim m_b)$ at the leading order in QED.  We have calculated
and confirm the results of Ref.~\cite{Bobeth:2003at} for all the
two-loop anomalous dimension matrices that determine the size of the
$\ln(M_H^2/M_L^2)$-enhanced electromagnetic corrections.

However, there are additional QED corrections that get enhanced by large
logarithms, namely $\ln(m_b^2/m_\ell^2)$. These corrections are the new
result of the present paper. They originate from these parts of the QED
bremsstrahlung corrections where the photon is collinear with one of the
outgoing leptons. They disappear after integration over the whole available
phase space but survive and remain numerically important when $m_{\ell\ell}^2$
is restricted to the region that we consider. 

 Such logarithmic corrections are found under the assumption that no
  collinear photons are included in the definition of the dilepton invariant
  mass.  This turns out to be a very good approximation for the muons in the
  current BaBar and Belle setups \cite{private}. In this case, the enhancement
  of the low-$m^2_{\ell\ell}$ integrated branching ratio by the collinear
  logarithms amounts to around 2\%. The corresponding effect for the electrons
  would reach around  5\%.   However, in that case, the logarithm of the electron
  mass gets replaced by the BaBar and Belle angular cut parameters and
  the integrated branching ratio for the electrons is expected to be close to
  that for the muons. We shall describe this issue in more detail in
  Section~\ref{sec:logdiscussion}. In the preceeding sections, our analytical
  and numerical results will correspond to the case of perfect separation of
  electrons and energetic collinear photons. 

Before we come to the results and details of the calculation, some
comments on its systematics are in order. Due to the different scales
involved, the perturbative corrections come not only with increasing
powers of some coupling constant, but also with increasing powers of
the large logarithm $L = \ln(M_H^2/M_L^2)$. The perturbative
calculation results in an expansion in the product
of the coupling with L rather than in the coupling alone.

Because $\alpha_s$ is relatively large, all powers of $c_s = \alpha_s L$ must be resummed at a
given order of the perturbative expansion, which is achieved using the
renormalization group technology. Within this framework, all the
logarithms $L$ are absorbed into $c_s = {\cal O}(1)$. Consequently,
each electromagnetic logarithm $\alpha_{\mathrm em} L = c_s
\alpha_{\mathrm em}/\alpha_s$ of the conventional perturbative
expansion gets replaced by $f(c_s) \alpha_{\mathrm em}/\alpha_s$,
where the function $f(c_s)$ is found by solving the renormalization
group equations.  Such a replacement of the electromagnetic logarithm
is not a matter of convenience but an unavoidable consequence of
resumming the QCD logarithms and not resumming the QED ones.
Resummation of the QED logarithms would be technically more difficult
and also unnecessary, because $\alpha_{\mathrm em} L \ll 1$.
Thus, the conventional expansion in $\alpha_s$ and
$\alpha_{\mathrm em}$ is replaced by an expansion in $\alpha_s$ and
in $\kappa\equiv \alpha_{\mathrm em}/\alpha_s$.  Each order of this
expansion is calculated exactly in $c_s$.

The amplitude of $B\to X_s \ell^+\ell^-$ is proportional to
$\alpha_{\mathrm em}$.  The Leading Order (LO) contributions come from
loops and are of order $\kappa$ (the electromagnetic logarithm comes
from a loop).  Higher order terms that are proportional to $\kappa
\alpha_s$ and $\kappa \alpha_s^2$ are conventionally called the NLO
and NNLO QCD contributions, respectively.  However, since $\kappa
\alpha_s = \alpha_{\mathrm em}$, the NLO contributions contain purely
electroweak terms, too. Since these NLO terms are enhanced by
$m_t^2/(M_W^2\sin^2\theta_W)$ while the LO terms are accidentally
suppressed, the two contributions turn out to be very similar in
size. An analogous effect occurs at order $\kappa^2$: the terms of
order $\kappa^2 \alpha_s^1$ are larger than the $\kappa^2 \alpha_s^0$
ones.  For this reason, also high terms in the $\kappa^n
\alpha_s^m$-expansion remain numerically important.

The corrections to be considered here (and also in
Ref~\cite{Bobeth:2003at}) are of order $\kappa^2$ and $\kappa^2
\alpha_s$ in the decay amplitude.  Contributions corresponding to
$\kappa^2 \alpha_s^2 \simeq \alpha_{\mathrm em}^2$ in the amplitude
will be included only if they are enhanced by $\ln(m_b^2/m_\ell^2)$
or by an additional factor of $m_t^2/(M_W^2\sin^2\theta_W)$.

Our article is organized as follows. In Section~\ref{sec:branchingratio} we
summarize the results for the branching ratio and explain details of the
$\kappa^n \alpha_s^m$-expansion. The effective theory used for resummation of
large QCD logarithms is introduced in Section \ref{sec:effectivetheory} which
is quite technical.  It includes the list of the relevant operators, the
matching conditions for the Wilson coefficients, the renormalization group
equations and the Wilson coefficients at the low scale.
Sections~\ref{sec:melem}~and~\ref{sec:emcorrp9} contain a detailed description
of the four-fermion operator matrix element calculation. In
Section~\ref{sec:logdiscussion} we discuss the  role of the angular
  cuts.  
Master formulae for the branching ratio are summarized in
Section~\ref{sec:masterBR}.  Appendix~\ref{sec:loopfunctions} contains the
loop functions that appear in the text. Some intermediate-step quantities for
the evolution of Wilson coefficients are collected in Appendix~\ref{app:Va}.
Appendix~\ref{app:details} is devoted to describing techniques that we have
used to calculate the QED matrix elements of quark-lepton operators.

\section{Branching ratio and numerical results}
\label{sec:branchingratio}
In order to facilitate the reading of this rather technical paper, we
give the final results first. The differential (with respect to $\s =
m^2_{\ell\ell}/m_{b,{\rm pole}}^2$) decay width of $\bar B \to X_s
\ell^+\ell^-$ can be expressed as follows:
\bea
\frac{{\rm d} \Gamma (\bar B \to X_s \ell^+\ell^-)}{{\rm d} \s} 
& = &
\frac{G_F^2 m_{b,{\rm pole}}^5}{48 \pi^3} \left| V_{ts}^* V_{tb} \right|^2 \; \Phi_{\ell\ell}(\s),
\eea
where the dimensionless function $\Phi_{\ell\ell}(\s)$ is assumed to
include both the perturbative and non-perturbative contributions.

In order to minimize the uncertainty stemming from $m_{b,{\rm pole}}^5$ and
the CKM angles, we normalize the rare decay rate to the measured semileptonic
one.  Furthermore, to avoid introduction of spurious uncertainties due to the
perturbative $b \to X_c e \bar{\nu}$ phase-space factor, we follow the 
  analyses of Refs.~\cite{Bobeth:2003at,Gambino:2001ew} where
\be
C = \left| \frac{V_{ub}}{V_{cb}} \right|^2 
       \frac{\Gamma (\bar B\to X_c e\bar\nu)}{\Gamma (\bar B\to X_u e\bar\nu)} \;,
\ee
was used instead. Consequently, our expression for the $\bar B \to X_s
\ell^+\ell^-$ branching ratio reads
\bea
{{\rm d} {\cal B} (\bar B\to X_s \ell^+\ell^-) \over {\rm d} \hat s} & = &
{\cal B} (B\to X_c e \bar\nu)_{\rm exp} \; 
\left| \frac{V_{ts}^* V_{tb}}{V_{cb}} \right|^2 \; 
\frac{4}{C} \; \frac{\Phi_{\ell\ell}(\s)}{\Phi_u} \;, 
\label{br}
\eea
where $\Phi_u = 1 + {\cal O}(\alpha_s,\alpha_{\mathrm em},\Lambda^2/m_b^2)$ is
defined by
\be \label{bu}
\Gamma (B\to X_u e\bar\nu) =
\frac{G_F^2 m_{b,{\rm pole}}^5}{192 \pi^3} \left| V_{ub}^{}\right|^2 \; \Phi_u.
\ee
Our expressions for the ratio $\Phi_{\ell\ell}(\s)/\Phi_u$ are
summarized in Section~\ref{sec:masterBR}. Both the perturbative and
non-perturbative corrections to this ratio are much better behaved
than for $\Phi_{\ell\ell}(\s)$ and $\Phi_u$ separately. The factor $C
= 0.58 \pm 0.01$ has been recently determined from a global analysis
of the semileptonic data \cite{Bauer:2004ve}. All the input parameters
that we use in the numerical calculation are summarized in
Table~\ref{tab:inputs}.
\begin{table}[t]
\begin{center}
\begin{displaymath}
\begin{tabular}{|l|l|}
\hline
\spp $\alpha_s (M_z) = 0.1182 \pm 0.0027$ \cite{Bethke:2004uy}
  &  $m_e = 0.51099892 \;\mev $ \\
\spp $\alpha_e (M_z) =  1/127.918 $ 
   & $m_\mu = 105.658369 \;\mev$ \\
\spp $s_W^2 \equiv \sin^2\theta_W = 0.2312$ 
   & $m_\tau = 1.77699 \;\gev$ \\
\spp $|V_{ts}^* V_{tb}/V_{cb}|^2 = 0.967 \pm 0.009$ \cite{Charles:2004jd}
   &  $m_c(m_c) = (1.224 \pm 0.017 \pm 0.054)\;\gev$ \cite{Hoang:2005zw}  \\
\spp $BR(B\to X_c e \bar\nu)_{\rm exp} = 0.1061 \pm 0.0017$ \cite{Aubert:2004aw}
   & $m_b^{1S} = (4.68 \pm 0.03)\;\gev$ \cite{Bauer:2004ve} \\
\spp $M_Z = 91.1876\;\gev$ 
   & $m_{t,{\rm pole}}= (172.7 \pm 2.9) \;\gev$ \cite{topmass}\\
\spp $M_W = 80.426\;\gev$
   & $m_B = 5.2794\;\gev$ \\
\spp $\lambda_2 \simeq \f{1}{4} \left(m_{B^*}^2-m_B^2\right) \simeq 0.12 \;\gev^2$ 
   & $C = 0.58 \pm 0.01$ \cite{Bauer:2004ve} \\\hline
\end{tabular}
\end{displaymath}
\caption{Numerical inputs that we use in the phenomenological
  analysis. Unless explicitly specified, they are taken from PDG 2004
  \cite{Eidelman:2004wy}.\label{tab:inputs}} 
\end{center}
\end{table}

 It should be stressed that the pole mass of the $b$ quark that is present in
the definition of $\s$ and in several loop functions gets analytically
converted to the so-called $1S$-mass before any numerical evaluation of the
branching ratio is performed. This way one avoids dealing with the renormalon
ambiguities in the definition of the pole mass \cite{Hoang:1998hm}. The
formula that relates the pole mass to the $1S$-mass can be found e.g. in
section 4 of Ref.~\cite{Hoang:2000fm}. 


Let us explain the details of the $\alpha_s$ and $\kappa$ expansion
that we adopt for calculating our final numerical results.  The $b\to
s \ell^+\ell^-$ decay amplitude has the following structure (up to an
overall factor of $G_F$):
\bea
{\cal A} & = & \hspace{5mm} \kappa \left[ {\cal A}_{LO}  + \alpha_s  \; {\cal A}_{NLO}+  
               \alpha_s^2 \; {\cal A}_{NNLO} + {\cal O}(\alpha_s^3) \right]
               \nonumber\\
         &   & +\kappa^2 \left[
               {\cal A}_{LO}^{\mathrm em} + \alpha_s \; {\cal A}_{NLO}^{\mathrm em}  + 
               \alpha_s^2  \; {\cal A}_{NNLO}^{\mathrm em} + {\cal O}(\alpha_s^3) \right] 
\; + {\cal O}(\kappa^3) \;.
\eea
As mentioned in the introduction, 
${\cal A}_{LO} \sim  \alpha_s  \; {\cal  A}_{NLO}$
and
${\cal A}_{LO}^{\mathrm em} \sim \alpha_s \; {\cal A}_{NLO}^{\mathrm em}$.
All these terms are included in our calculation in a complete manner, together
with the appropriate bremsstrahlung corrections. As far as ${\cal A}_{NNLO}$
is concerned, we use the practically complete results of
Refs.~\cite{Bobeth:1999mk, Asatryan:2001zw, Asatryan:2002iy, Ghinculov:2003qd,
  Gambino:2003zm, Gorbahn:2004my ,Bobeth:2003at}; the only missing parts
originate from the unknown two-loop matrix elements of the QCD-penguin
operators whose Wilson coefficients are very small.

Among the contributions to ${\cal A}_{NNLO}^{\mathrm em}$, we include
only the terms which are either enhanced by an additional factor of
$m_t^2/(M_W^2\sin^2\theta_W)$ (with respect to ${\cal
A}_{NLO}^{\mathrm em}$) \cite{Bobeth:2003at} or contribute to the $\ln
(m_b^2/m_\ell^2)$-enhanced terms at the decay width level. The latter
terms are calculated for the first time here. They are taken into
account in a practically complete manner; the only missing part is
proportional to the same tiny Wilson coefficient that is responsible
for the smallness of ${\cal A}_{LO}$.

The perturbative expansion of the ratio $\Phi_{\ell\ell}(\s)/\Phi_u$
has  a similar  structure  to that of the squared amplitude:
\bea
{\cal A}^2 
& = & 
\kappa^2 \Big[ {\cal A}_{LO}^2 
+  \alpha_s \; 2 {\cal A}_{LO} {\cal A}_{NLO} 
+ \alpha_s^2 \; ({\cal A}_{NLO}^2 + 2  {\cal A}_{LO} {\cal A}_{NNLO}  ) 
\nonumber\\
& & \hskip 0.5 cm 
+  \alpha_s^3 \; 2 ({\cal A}_{NLO} {\cal A}_{NNLO} + \ldots)  + {\cal O}(\alpha_s^4))
\Big]   \nonumber \\
&+ &
\kappa^3 \Big[
2 {\cal A}_{LO}^{} {\cal A}_{LO}^{\mathrm em} + 
\alpha_s \; 2 ({\cal A}_{NLO}^{} {\cal A}_{LO}^{\mathrm em} + {\cal A}_{LO}^{} {\cal A}_{NLO}^{\mathrm em}) 
\nonumber\\
& &
\hskip 0.5 cm 
+\alpha_s^2 \; 2 ( {\cal A}_{NLO}^{} {\cal A}_{NLO}^{\mathrm em}
+ {\cal A}_{NNLO}^{} {\cal A}_{LO}^{\mathrm em} 
+ {\cal A}_{LO}^{} {\cal A}_{NNLO}^{\mathrm em}) \nonumber\\
&& \hskip 0.5 cm
+ \alpha_s^3 \; 2 ({\cal A}_{NLO}^{} {\cal A}_{NNLO}^{\mathrm em} + {\cal A}_{NNLO}^{} {\cal A}_{NLO}^{\mathrm em}
+ \ldots)
+ {\cal O}(\alpha_s^4)\Big] \nonumber \\
&+ &
 {\cal O}( \kappa^4)\; .
\eea
In our numerical calculation of $\Phi_{\ell\ell}(\s)/\Phi_u$, we include all
the terms that are written explicitly in the above equations. The dots at
orders $\kappa^2 \alpha_s^3$ and $\kappa^3 \alpha_s^3$ stand for terms that
are proportional to ${\cal A}_{LO}$ and ${\cal A}_{LO}^{\mathrm em}$ and,
consequently, can safely be neglected. In the numerical analysis we also
include subleading $1/m_c$ and $1/m_b$
corrections~\cite{Buchalla:1997ky,Buchalla:1998mt, Falk:1993dh} as well as
finite bremsstrahlung effects~\cite{Asatryan:2002iy}.

Our results for the branching ratios integrated in the range
$1\;\gev^2 < m^2_{\ell\ell} < 6\;\gev^2$ read
 \bea
\hskip -0.5cm
{\cal B}_{\mu\mu} & = & \label{muonBR} 
\Big[ 
 1.59 
\pm 0.080_{\rm scale}
\pm 0.06_{m_t} 
\pm 0.026_{ C,m_c }
\pm 0.015_{m_b} \nnb \\
&& \hspace*{28.5pt} \pm 0.02_{\alpha_s(M_Z)}
\pm 0.015_{\rm CKM} 
\pm 0.025_{{\rm BR}_{sl}} 
\Big] \times 10^{-6} = (  1.59  \pm 0.11 ) \times 10^{-6} \;, \\
\hskip -0.5cm
{\cal B}_{ee} & = & \label{electronBR} 
\Big[ 
 1.64  
\pm 0.085_{\rm scale} 
\pm 0.06_{m_t} 
\pm 0.027_{ C,m_c }
\pm 0.015_{m_b} \nnb \\
&& \hspace*{28.5pt} \pm 0.02_{\alpha_s(M_Z)}
\pm 0.015_{\rm CKM}
\pm 0.026_{{\rm BR}_{sl}}
\Big] \times 10^{-6}  = (  1.64   \pm 0.11) \times 10^{-6} \;.
\eea 
The central values are obtained for the matching scale $\mu_0 = 120\;\gev$ and
the low-energy scale $\mu_b=5\;\gev$. The uncertainty from missing higher
order perturbative corrections have been estimated by increasing and
decreasing the scales $\mu_{0,b}$ by factors of 2. Uncertainties induced by
$m_t$, $m_b$, $m_c$, $C$, $\alpha_s(M_Z)$, the CKM angles and the semileptonic
rate are obtained by varying the various inputs within the errors given in
Table~\ref{tab:inputs}; we assume the errors on C and $m_c$ to be  fully 
correlated. The total error is obtained by adding the individual uncertainties
in quadrature.  The electron and muon channels receive different contributions
because of the $\ln(m_b^2/m_\ell^2)$ present in the bremsstrahlung
corrections.  The difference gets reduced when the BaBar and Belle
angular cuts are included (see Sec.~\ref{sec:logdiscussion}). 

We stress that the indicated uncertainties are only the parametric and
perturbative ones. No additional uncertainty for the unknown subleading
non-perturbative corrections has been included. In particular, we believe that
the uncalculated order $\alpha_s(\mu_b) \f{\Lambda_{QCD}}{m_{c,b}}$
non-perturbative corrections imply an additional uncertainty of around  
$\sim 5\%$ in the above formula. This issue deserves an independent study.

One should also keep in mind that all the effects of the intermediate $\psi$
and $\psi'$ contributions are assumed to be subtracted on the experimental
side. This refers, in particular, to the decays $\psi \to X \ell^+ \ell^-$
where low-mass dilepton pairs can be produced. All such decays of the $\psi$
with branching ratios down to $10^{-5}$ may be relevant. To our knowledge, only
$X = \gamma$ has been considered so far in the experimental analyses. 

 The overall uncertainties in Eqs.~(\ref{muonBR}) and (\ref{electronBR})
  are somewhat smaller than in Eq.~(27) of Ref.~\cite{Bobeth:2003at}. This is
  mainly due to the improved experimental value of $m_t$ as well as to our use of  
  $m_b^{1S}$ rather than $m_{b,{\rm pole}}$. The latter opportunity was already
  suggested in Ref.~\cite{Bobeth:2003at}.
\begin{table}[t]
 \begin{center}
\begin{tabular}{|l|l||l|l|}
\hline
 \spp  NLO ($\alpha_{\mathrm em} (\mu_0)$) & $ 1.81 \times 10^{-6}$   &
 \spp  NLO ($\alpha_{\mathrm em} (\mu_b)$) & $ 1.69 \times 10^{-6}$  \\ \hline
 \spp  NNLO ($\alpha_{\mathrm em} (\mu_0)$) & $ 1.65\times 10^{-6}$  &
 \spp  NNLO ($\alpha_{\mathrm em} (\mu_b)$) & $ 1.53 \times 10^{-6}$   \\ \hline \hline
 \spp  QED (only WC's) & $ 1.56\times 10^{-6}$  & & \\  \hline 
 \spp  QED (muons) & $ 1.59\times 10^{-6}$        &
 \spp  QED (electrons) &  $ 1.64\times 10^{-6}$  \\ \hline
\end{tabular}
\caption{Anatomy of QCD and QED corrections.  \label{tab:results}}
\end{center} 
\end{table} 

In Table~\ref{tab:results}, we show the partial results that we obtain
by adding sequentially all the known QCD and QED corrections. The rows
denoted by ``NLO'' and ``NNLO'' refer to the leading order in QED. The
row ``QED (only WC's)'' contains only those QED corrections that stem
from the Wilson coefficients. The row ``QED'' includes all the
electromagnetic corrections (that are different for electrons and
muons,  as in Eqs.~(\ref{muonBR}) and (\ref{electronBR}) ).

A numerical formula that gives the branching ratio for non-SM values of the
high-scale Wilson coefficients of the operators $P_7$, $P_8$, $P_9$ and
$P_{10}$ (see Section~\ref{sec:effectivetheory}) reads

\bea\label{numform} 
\hskip -3cm 
{\cal B}_{\mu\mu} & = & \Big[\; 
2.1774 - 0.001658 \; {\cal I} (R_{10})+ 0.0005 \; { \cal I}( R_{10} 
R_{8}^*) + 0.0534 \; {\cal I} (R_{7})+ 0.02266 \; { \cal I} (R_{7} 
R_{8}^*)
\nonumber\\ & & 
+ 0.00496\; { \cal I}( R_{7} R_{9}^*)+ 0.00527 \; {\cal I} 
(R_{8})+ 0.0261 \; {\cal I} (R_{8} R_{9}^*)- 0.0115 \; {\cal I} (R_{9})
\nonumber\\ & & 
- 0.5420 \; {\cal R} (R_{10})+ 0.0208 \; {\cal R} (R_{7})+ 0.0153 \; { \cal R}( R_{7} 
R_{10}^*)+ 0.06848 \; { \cal R} (R_{7} R_{8}^*)
\nonumber\\ & & 
- 0.8545 \; { \cal R} (R_{7} 
R_{9}^*)- 0.00938 \; {\cal R} (R_{8}) + 0.00185 \; { \cal R}( R_{8} 
R_{10}^*)- 0.0981 \; { \cal R} (R_{8} R_{9}^*)
\nonumber\\ & & 
+ 2.6917 \; {\cal R} 
(R_{9})- 0.10698 \; {\cal R}( R_{9} R_{10}^*)+ 10.7652 \; |R_{10}|^2+ 0.2880 \; 
|R_7|^2
\nonumber\\ & & 
+ 0.00381 \; |R_8|^2 + 1.4884 \; |R_9|^2\; \Big] \times 10^{-7} \; ,\\
{\cal B}_{ee} & = & \Big[\; 
2.3148 - 0.001658 \; {\cal I} (R_{10})+ 0.0005 \; { \cal I}( R_{10} \
R_{8}^*)+ 0.0523 \; {\cal I} (R_{7})+  0.02266 \; { \cal I} (R_{7} \
R_{8}^*)
\nonumber\\ & & 
+ 0.00496 \; { \cal I}( R_{7} R_{9}^*)+ 0.00518 \; {\cal I} \
(R_{8})+ 0.0261\; {\cal I} (R_{8} R_{9}^*) - 0.00621 \; {\cal I} (R_{9})
\nonumber\\ & & 
- 0.5420 \; {\cal R} (R_{10}) - 0.03340 \; {\cal R} (R_{7}) + 0.0153 \; { \cal R}( R_{7} \
R_{10}^*)+ 0.0673 \; { \cal R} (R_{7} R_{8}^*)
\nonumber\\ & & 
- 0.86916 \; { \cal R} (R_{7} \
R_{9}^*) - 0.0135 \; {\cal R} (R_{8})+ 0.00185 \; { \cal R}( R_{8} \
R_{10}^*)- 0.09921 \; { \cal R} (R_{8} R_{9}^*)
\nonumber\\ & & 
 + 2.833 \; {\cal R} \
(R_{9})- 0.10698 \; {\cal R}( R_{9} R_{10}^*)+ 11.0348 \; |R_{10}|^2+ 0.2804 \; \
|R_7|^2
\nonumber\\ & & 
+ 0.003763 \; |R_8|^2+ 1.527 \; |R_9|^2
\; \Big] \times 10^{-7} \; , 
\eea 
where  (see Eqs.~(\ref{matching1}), (\ref{c7eff}), (\ref{c8eff}) for the
  Wilson coefficient definitions)  
\bea
R_{7,8} = \frac{C_{7,8}^{(00){\rm eff}} (\mu_0)}{C_{7,8}^{(00){\rm eff,SM}}(\mu_0)}  
\hspace{1cm} {\rm and} \hspace{1cm}
R_{9,10} = \frac{C_{9,10}^{(11)} (\mu_0)}{C_{9,10}^{(11){\rm SM}}(\mu_0)} \;.
\eea

\section{The effective theory }
\label{sec:effectivetheory}
\subsection{Operator basis \label{sec:operators}}

Resummation of large QCD logarithms is most conveniently performed in
the framework of a low-energy effective theory~\cite{Buras:1998ra}. There are
ten operators that need to be considered at the leading order in the
electroweak interactions. They can be chosen as follows:
\bea \label{ope}
\begin{array}{rl}
P_1   = & (\bar{s}_L \gamma_{\mu} T^a c_L) (\bar{c}_L \gamma^{\mu} T^a b_L),
\vspace{0.2cm} \\
P_2   = & (\bar{s}_L \gamma_{\mu}     c_L) (\bar{c}_L \gamma^{\mu}     b_L),
\vspace{0.2cm} \\
P_3   = & (\bar{s}_L \gamma_{\mu}     b_L) \sum_q (\bar{q}\gamma^{\mu}     q),
\vspace{0.2cm} \\
P_4   = & (\bar{s}_L \gamma_{\mu} T^a b_L) \sum_q (\bar{q}\gamma^{\mu} T^a q),    
\vspace{0.2cm} \\
P_5   = & (\bar{s}_L \gamma_{\mu_1}
                     \gamma_{\mu_2}
                     \gamma_{\mu_3}    b_L)\sum_q (\bar{q} \gamma^{\mu_1} 
                                                         \gamma^{\mu_2}
                                                         \gamma^{\mu_3}     q),     
\vspace{0.2cm} \\
P_6   = & (\bar{s}_L \gamma_{\mu_1}
                     \gamma_{\mu_2}
                     \gamma_{\mu_3} T^a b_L)\sum_q (\bar{q} \gamma^{\mu_1} 
                                                            \gamma^{\mu_2}
                                                            \gamma^{\mu_3} T^a q),
\vspace{0.2cm} \\
P_7   = &  \f{e}{16 \pi^2} m_b (\bar{s}_L \sigma^{\mu \nu}     b_R) F_{\mu \nu},
\vspace{0.2cm} \\
P_8   = &  \f{g}{16 \pi^2} m_b (\bar{s}_L \sigma^{\mu \nu} T^a b_R) G_{\mu \nu}^a, 
\vspace{0.2cm} \\
P_9      = & (\bar{s}_L \gamma_{\mu} b_L) \sum_l (\bar{l}\gamma^{\mu} l),
\vspace{0.2cm} \\
P_{10}   = & (\bar{s}_L \gamma_{\mu}     b_L) \sum_l (\bar{l}\gamma^{\mu} \gamma_5 l).
\end{array} 
\eea
In $P_3$, ..., $P_6$, the quark flavors are $q = u,d,s,c,b$. In $P_9$ and
$P_{10}$, all the three lepton flavors are present.  Contrary to other
analyses \cite{Misiak:1992bc,Buras:1994dj},  we have not included any gauge
couplings in the normalization of $P_9$ and $P_{10}$. Including them would
give only a minor simplification in the present investigation.

Once QED corrections are considered, five more operators
need to be taken into account. They can be chosen as
\bea
\begin{array}{rl}
P_{3Q} = & (\bar{s}_L \gamma_{\mu}     b_L) \sum_q Q_q (\bar{q}\gamma^{\mu}     q),    
\vspace{0.2cm} \\
P_{4Q} = & (\bar{s}_L \gamma_{\mu} T^a b_L) \sum_q Q_q (\bar{q}\gamma^{\mu} T^a q),   
\vspace{0.2cm} \\
P_{5Q} = & (\bar{s}_L \gamma_{\mu_1}
                     \gamma_{\mu_2}
                     \gamma_{\mu_3}    b_L)\sum_q Q_q (\bar{q} \gamma^{\mu_1} 
                                                               \gamma^{\mu_2}
                                                               \gamma^{\mu_3}     q),
\vspace{0.2cm} \\
P_{6Q} = & (\bar{s}_L \gamma_{\mu_1}
                     \gamma_{\mu_2}
                     \gamma_{\mu_3} T^a b_L)\sum_q Q_q (\bar{q} \gamma^{\mu_1} 
                                                                \gamma^{\mu_2}
                                                                \gamma^{\mu_3} T^a q),     
\vspace{0.2cm} \\
P_b = & \f{1}{12} \left[ 
          (\bar{s}_L \gamma_{\mu_1}
                     \gamma_{\mu_2}
                     \gamma_{\mu_3}    b_L)            (\bar{b} \gamma^{\mu_1} 
                                                                \gamma^{\mu_2}
                                                                \gamma^{\mu_3}     b)
             -4 (\bar{s}_L \gamma_{\mu} b_L) (\bar{b} \gamma^{\mu} b) \right].
\end{array} 
\eea
where $Q_q$ are the electric charges of the corresponding quarks
($\f{2}{3}$ or $-\f{1}{3}$).

The Lagrangian of the effective theory reads
\mathindent0cm
\be \label{Leff}
{\cal L}_{eff} = {\cal L}_{\scs QCD \times QED}(u,d,s,c,b,e,\mu,\tau)  
+\f{4 G_F}{\sqrt{2}} V^*_{ts} V_{tb} 
\left[ \sum_{i=1}^{10} C_i(\mu) P_i + \sum_{i=3}^{6} C_{iQ}(\mu) P_{iQ} + C_b(\mu) P_b \right].
\ee
\mathindent1cm

\subsection{Matching conditions \label{sec:matching}}

The Wilson coefficients at the matching scale $\mu_0 \sim M_W, m_t$ are
expanded as follows
\bea \label{matching1}
C_k(\mu_0) &=&   C_k^{(00)}(\mu_0) 
+ \as( \mu_0) \; C_k^{(10)}(\mu_0)  
+ \as(\mu_0)^2 \; C_k^{(20)}(\mu_0) \nonumber\\ 
& &  
+ \as(\mu_0) \kappa(\mu_0) \; C_k^{(11)}(\mu_0) 
+ \as(\mu_0)^2 \kappa(\mu_0)\;  C_k^{(21)}(\mu_0) 
+ {\cal O}(\as^3,\kappa^2\as^2),
\eea
where $\widetilde{\alpha}_s = \alpha_s/4\pi$. Note, that at the low scale
$\mu_b \sim m_b, m_{\ell\ell}$, also terms of order $\kappa$, $\kappa^2$ and
$\kappa^2 \alpha_s$ arise and are included wherever necessary.

The values of the Wilson coefficients are found from the requirement that all
the effective theory Green functions\footnote{
  For the on-shell 1PR functions, the operators from Section
  \ref{sec:operators} are sufficient. However, it is often more
  convenient find the Wilson coefficients by matching the off-shell
  1PI functions. Then, additional operators are necessary --- see
  Eq.~(73) of Ref.~\cite{Bobeth:1999mk}.}
match to the full SM ones at the leading order in $M_L^2/M_H^2$.  At the order
we consider, the following non-vanishing contributions to
Eq.~(\ref{matching1}) must be taken into account for the four-fermion
operators  ($s_W^2 \equiv \sin^2\theta_W$): 
\bea
C_2^{(00)}(\mu_0) &=& 1\;,\\
C_1^{(10)}(\mu_0) &=& 15 + 6 \ln\f{\mu_0^2}{M_W^2}\;,\\
C_4^{(10)}(\mu_0) &=& E(x_t) - \f{2}{3} + \f{2}{3} \ln\f{\mu_0^2}{M_W^2}\;,\\
C_2^{(11)}(\mu_0) &=& -\f{7}{3} - \f{4}{3} \ln\f{\mu_0^2}{M_Z^2}\;,\\
C_3^{(11)}(\mu_0) &=& \f{2}{9s_W^2} \left[ X(x_t) - 2Y(x_t) \right]\;,\\
C_5^{(11)}(\mu_0) &=& -\f{1}{18s_W^2} \left[ X(x_t) - 2Y(x_t) \right],\\
C_9^{(11)}(\mu_0) &=& \f{1}{s_W^2} Y(x_t) + W(x_t) + \f{4}{9} 
                     - \f{4}{9} \ln\f{\mu_0^2}{m_t^2}\;,\\
C_{10}^{(11)}(\mu_0) &=& -\f{1}{s_W^2} Y(x_t)\;,\\
C_{3Q}^{(11)}(\mu_0) &=& \f{2}{3s_W^2} \left[ X(x_t) + Y(x_t) \right]
                 - W(x_t) - \f{4}{9} + \f{4}{9} \ln\f{\mu_0^2}{m_t^2}\;,\\
C_{5Q}^{(11)}(\mu_0) &=& -\f{1}{6s_W^2} \left[ X(x_t) + Y(x_t) \right]\;,\\
C_b^{(11)}(\mu_0) &=& -\f{1}{2s_W^2} S(x_t)\;, \\
C_i^{(20)}(\mu_0) &=&  C_i^{t(2)}(\mu_0) - C_i^{c(2)}(\mu_0) 
                        \;\;\; {\rm for} \;\; i=1,...,6 \;,\\
C_i^{(21)}(\mu_0) &=&  C_i^{t(2)}(\mu_0) - C_i^{c(2)}(\mu_0) 
                        \;\;\; {\rm for} \;\; i=9,10 \;,\\
C_9^{(22)}(\mu_0) &=&  - {x_t^2 \over 32 s_W^4} \; (4 s_W^2 -1) \; \left[ 3 + \tau_b^{(2)}( x_{ht})-
                         \Delta_t (x_{ht}) \right]\;,\\
C_{10}^{(22)}(\mu_0) &=&  - {x_t^2 \over 32 s_W^4} \left[ 3 + \tau_b^{(2)}( x_{ht})-
                         \Delta_t (x_{ht}) \right]\;.
\eea
All the one-loop coefficients $C_i^{(1m)}(\mu_0)$ above have been
evaluated in the ${\overline{\rm MS}}$ scheme.\footnote{
  Beyond tree-level, the Wilson coefficients usually depend on the
  choice of evanescent operators.  Our choice is the same as in 
  Refs.~\cite{Gambino:2003zm,Gorbahn:2004my,Bobeth:2003at}. }
The functions $E(x)$, $X(x)$, $Y(x)$, $W(x)$, $S(x)$ are collected in
Appendix~\ref{sec:loopfunctions}.  The one-loop coefficient $C_2^{(11)}$ is
from Ref.~\cite{Gambino:2001au}. The other one-loop ones have been known since
many years (see, e.g., Ref.~\cite{Buchalla:1989we}).  For $C_i^{(20)}$ and
$C_i^{(21)}$, the relevant top ($C_i^{t(2)}$) and charm ($C_i^{c(2)}$)
contributions to the two-loop matching conditions can be found in Section 2 of
Ref.~\cite{Bobeth:1999mk}.  The functions $\tau_b^{(2)}$ and $\Delta_t$,
  where $x_t\equiv (m_t^{\overline {MS}}/M_W)^2$ and $x_{ht} \equiv
  (M_h/m_t^{\overline {MS}})^2$, can be found in Ref.~\cite{Buchalla:1997kz}.
  We include also the contributions  to $C_{i(Q)}^{(21)} (\mu_0)$
that were calculated in Refs.~\cite{Bobeth:1999mk,Buras:1999st}.  
Transforming the results of Ref.~\cite{Buras:1999st} to our operator basis
is non-trivial.\footnote{
We thank  Ulrich  Haisch for providing us with the  relevant
transformation matrices.}

\subsection{Renormalization Group Equations \label{sec:RGE}}

In the effective theory, the RGE for the gauge couplings read 
\bea
\mu \f{d\as}{d\mu}  &=& - 2 \as^2 \sum_{n,m=0} \beta_{nm}^s \as^n \aem^m,
\nnb\\[-2mm] && \label{gaugeRGE} \\[-3mm]
\mu \f{d\aem}{d\mu} &=& + 2 \aem^2 \sum_{n,m=0} \beta_{nm}^e \aem^n \as^m,
\nnb
\eea
where $\widetilde{\alpha}_e = \alpha_{\rm em}/4\pi$.
The solution for $\as(\mu)$ with the initial condition at $\mu = \mu_0$ is
found perturbatively in $\as(\mu_0)$ and $\aem(\mu_0)$ but exactly in~~ $v_s =
1 \!+\! 2 \beta_{00}^s \as(\mu_0) \ln \f{\mu}{\mu_0}$~~ and ~~$v_e = 1 \!-\! 2
\beta_{00}^e \aem(\mu_0) \ln \f{\mu}{\mu_0}$.~~ Including all the 3-loop
contributions, and, in addition, the 4-loop pure-QCD term, one obtains
\mathindent0cm
\bea
\as(\mu) &=& \f{\as(\mu_0)}{v_s} ~-~ \f{\as(\mu_0)^2}{v_s^2} 
  \left( \f{\beta_{10}^s}{\beta_{00}^s} \ln v_s - \f{\beta_{01}^s}{\beta_{00}^e} \ln v_e \right) 
   ~+~ \f{\as(\mu_0)^3}{v_s^3} \left[ \;\f{\beta_{20}^s}{\beta_{00}^s} (1-v_s) 
\right. \nnb \\[2mm] &+& \left. \hspace{-1mm}  
\left(\f{\beta_{10}^s}{\beta_{00}^s}\right)^2 
\left( \ln^2 v_s-\ln v_s + v_s - 1 \right)
+ \left(\f{\beta_{01}^s}{\beta_{00}^e}\right)^2 \ln^2 v_e
+ \f{\beta_{01}^s\beta_{10}^s}{\beta_{00}^s\beta_{00}^e} 
                      \left( -2 \ln v_s \ln v_e + \rho \, v_e \ln v_e \right) \right]
\nnb \\[2mm] &+& 
\f{\as(\mu_0)^4}{v_s^4} \left[ \;\f{\beta_{30}^s}{\beta_{00}^s} \f{1-v_s^2}{2} 
+ \f{\beta_{20}^s\beta_{10}^s}{(\beta_{00}^s)^2} \left( (2v_s-3) \ln v_s +v_s^2-v_s \right)
\right. \nnb \\[2mm] &+& \left. 
\left(\f{\beta_{10}^s}{\beta_{00}^s}\right)^3 
\left( -\ln^3 v_s + \f{5}{2} \ln^2 v_s + 2(1-v_s) \ln v_s - \f{1}{2} (v_s - 1)^2 \right) \right]
\nnb \\[2mm] &+& 
\f{\as(\mu_0)^2 \aem(\mu_0)}{v_s^2 v_e} \left[ \;\f{\beta_{02}^s}{\beta_{00}^e} (v_e-1) 
+ \f{\beta_{11}^s}{\beta_{00}^s} \rho \, v_e \ln \f{v_e}{v_s}
+ \f{\beta_{01}^s\beta_{10}^e}{(\beta_{00}^e)^2} \left( \ln v_e - v_e + 1 \right)
\right. \nnb \\[2mm] &+& \left. 
\f{\beta_{01}^s\beta_{10}^s}{(\beta_{00}^s)^2} \rho \, v_e \ln v_s
+ \f{\beta_{01}^s\beta_{01}^e}{\beta_{00}^s\beta_{00}^e} 
                      \left( \rho \, v_e \ln \f{v_s}{v_e} - \ln v_s \right) 
\right] ~+~ \as^2 \times {\cal O}(\as^3, \aem^2, \as \aem),
\label{alpharun}
\eea
\mathindent1cm
where 
$\rho = \beta_{00}^s \as(\mu_0)/(\beta_{00}^s \as(\mu_0)+ \beta_{00}^e \aem(\mu_0))$.  
The corresponding solution for $\aem(\mu)$ can be found by obvious
replacements: $v_s \leftrightarrow v_e$,~ ~$\as \leftrightarrow \aem$~
and ~$\beta_{ij}^s \leftrightarrow -\beta_{ij}^e$ (also inside the
ratio $\rho$).

The $\overline{\rm MS}$ values of the pure-QCD coefficients
$\beta^s_{i0}$ can be found in Refs.~\cite{vanRitbergen:1997va,Czakon:2004bu}.  After
substituting $C_A = N = 3$, $C_F = \fm{4}{3}$, $t_F = \f{1}{2}$ and
$n_f = 5$, one finds $\beta_{00}^s = \f{23}{3}$, $\beta_{10}^s =
\f{116}{3}$, $\beta_{20}^s = \f{9769}{54}$ and $\beta_{30}^s =
\f{352864}{81}\zeta(3)-\f{598391}{1458}$. The remaining beta-function
coefficients that enter into Eq.~(\ref{alpharun}) read\footnote{
 All of them except $\beta_{11}^s$ can be found by modifying the color and
charge factors in the pure QCD results. As far as $\beta_{11}^s$ is concerned,
we have found it by performing an explicit three-loop calculation. To our
knowledge, no result for this coefficient has been published before. }
\be \label{betafuncRGE} 
\begin{array}{lcl}
\beta_{01}^s = -4 t_F \overline{Q^2} = -\f{22}{9}, &\hspace{1cm}&
\beta_{00}^e = \fm{4}{3} \left( \overline{Q^2} N + 3 Q_l^2 \right) = \fm{80}{9},\\[2mm]
\beta_{11}^s = \left(4 C_F -8 C_A\right) t_F \overline{Q^2} = -\fm{308}{27} &&
\beta_{10}^e = 4 \left( \overline{Q^4} N + 3 Q_l^4 \right) = \fm{464}{27},\\[2mm]
\beta_{02}^s = \fm{11}{3} t_F \overline{Q^2} \beta_{00}^e + 2 t_F \overline{Q^4} = \fm{4945}{243} &&
\beta_{01}^e =  4 C_F N \overline{Q^2} = \fm{176}{9},
\end{array}
\ee
where $Q_l = -1$, $Q_u = \fm{2}{3}$, $Q_d = -\fm{1}{3}$ and $\overline{Q^n}
= 2 Q_u^n + 3 Q_d^n$.\\

The RGE for the Wilson coefficients reads
\be \label{WilsonRGE}
\mu \f{d}{d\mu} \vec{C}(\mu) = \hat{\gamma}^T(\mu) \vec{C}(\mu), 
\ee
where the Anomalous Dimension Matrix (ADM) has the following expansion:
\be \label{ADMexp}
\hat{\gamma}(\mu) = \sum_{\begin{array}{c} \\[-7mm] \scriptstyle{n,m=0} \\[-2mm] 
   \scriptstyle{n+m\geq 1}\\[-3mm] \end{array}} \hat{\gamma}^{(nm)} \as(\mu)^n \aem(\mu)^m.
\ee

In Eq.~(\ref{alpharun}), we have made no use of the fact that $\aem
\ll \as$. Now we shall take this relation into account, and solve the
RGE (\ref{WilsonRGE}) perturbatively in
\be
\lambda \equiv \f{\beta_{00}^e \, \aem(\mu_0)}{\beta_{00}^s \, \as(\mu_0)} 
\hspace{2cm} {\rm and}  \hspace{2cm} 
\omega \equiv 2 \beta_{00}^s \as(\mu_0),
\ee
neglecting terms of order ${\cal O}(\omega^3, \lambda^3, \omega^2
\lambda^2)$.  Let us introduce the following short-hand notation:
\bea && \hspace{-7mm}
b_1 = \f{\beta_{10}^s}{2 (\beta_{00}^s)^2}, \hspace{12mm} 
b_2 = \f{\beta_{20}^s}{4 (\beta_{00}^s)^3} - b_1^2, \hspace{12mm} 
b_3 = \f{\beta_{01}^s}{2 \beta_{00}^s \beta_{00}^e}, \nnb\\ && \hspace{-7mm}
b_4 = \f{\beta_{11}^s}{4 (\beta_{00}^s)^2 \beta_{00}^e} - 2 b_1 b_3, \hspace{1cm} 
b_5 = \f{\beta_{01}^e}{2 \beta_{00}^s \beta_{00}^e} - b_1, \hspace{1cm} 
\hat{W}^{(nm)} = \f{\left(\hat{\gamma}^{(nm)}\right)^T}{(2 \beta_{00}^s)^n (2 \beta_{00}^e)^m}.
\eea
The known evolution of the gauge couplings (\ref{alpharun})
allows us to rewrite the RGE (\ref{WilsonRGE}) in terms of the variable
$\eta = \as(\mu_0)/\as(\mu)$
\be \label{etaRGE}
\f{d}{d\eta} \vec{C} = \f{1}{\eta} \left[ \hat{W}^{(10)} ~+~ \sum_{k=-2}^2 \hat{B}^{(k)} \eta^k
 ~+~ \lambda^2 \omega b_5 \hat{W}^{(01)} \eta \ln \eta 
 ~+~ {\cal O} (\omega^3,\lambda^3,\omega^2\lambda^2) \right] \vec{C}. 
\ee
where the matrices $\hat{B}^{(k)}$ are $\eta$-independent
\bea
\hat{B}^{(-2)}\!\!\! &=& \omega^2 \left( \hat{W}^{(30)}\! - b_1 \hat{W}^{(20)}\! - b_2 \hat{W}^{(10)} \right),\\
\hat{B}^{(-1)}\!\!\! &=& \omega \left( \hat{W}^{(20)}\! - b_1 \hat{W}^{(10)} \right)
                    + \omega^2 \lambda \left( \hat{W}^{(21)}\! - b_1 \hat{W}^{(11)}\! 
                    - b_2 \hat{W}^{(01)}\! -b_3 \hat{W}^{(20)}\! - b_4 \hat{W}^{(10)}\right),\\
\hat{B}^{( 0)} &=&  \omega \lambda (1-\lambda) 
                  \left( \hat{W}^{(11)}\! - b_1 \hat{W}^{(01)}\! - b_3 \hat{W}^{(10)} \right),\\
\hat{B}^{( 1)} &=&  \lambda ( 1-\lambda) \hat{W}^{(01)} + \omega \lambda^2 \left( \hat{W}^{(02)}\! 
                     + \hat{W}^{(11)}\! -(b_1+b_3) \hat{W}^{(01)}\! - b_3 \hat{W}^{(10)} \right),\\ 
\hat{B}^{(2)} &=& \lambda^2 \hat{W}^{(01)}.
\eea

The solution to Eq.~(\ref{etaRGE}) reads 
\bea
\vec{C}(\mu)\! &=& \!\hat{V} \left[ \hat{D}(\eta) 
+\hspace{-2mm} \sum_{k=-2}^2 \hat{F}^{(k)}(\eta) 
+\hspace{-2mm} \sum_{k,l=-2}^2 \! \hat{G}^{(kl)}(\eta) 
\right. \nnb \\ && \hspace{25mm} \left.
+\hspace{-2mm} \sum_{k,l,m=-2}^2 \!\! \hat{H}^{(klm)}(\eta) 
+ \hat{R}(\eta) + {\cal O}(\omega^3,\lambda^3,\omega^2\lambda^2)
\right] \hat{V}^{-1} \vec{C}(\mu_0),
\label{cimb}
\eea
where $\hat{V}$ is the matrix that diagonalizes $\hat{W}^{(10)}$
\be
\left[ \hat{V}^{-1} \hat{W}^{(10)} \hat{V} \right]_{ij} = a_{\underline{i}} \delta_{ij}.
\ee
The eigenvalues $a_i$ and entries of the matrix $\hat{V}$ are given
numerically in appendix~\ref{app:Va}.  The matrices $\hat{D}(\eta)$,
$\hat{F}^{(k)}(\eta)$, $\hat{G}^{(kl)}(\eta)$, $\hat{H}^{(klm)}(\eta)$
and $\hat{R}(\eta)$ depend on the $a_i$ and on products $\hat{E}^{(k)}
\equiv \hat{V}^{-1} \hat{B}^{(k)} \hat{V}$ . They read
\bea && \hspace{-7mm}
\hat{D}_{ij}(\eta) = \eta^{a_{\underline{i}}} \delta_{ij}, \hspace{1cm}
\hat{F}^{(k)}_{ij}(\eta) = \hat{E}^{(k)}_{ij} f^{(k)}_{\underline{i}\underline{j}}(\eta), \hspace{1cm}
\hat{G}^{(kl)}_{ij}(\eta) = \sum_p \hat{E}^{(k)}_{ip} \hat{E}^{(l)}_{pj} 
                       g^{(kl)}_{\underline{i} p \underline{j}}(\eta),
\nnb \\[2mm] && \hspace{-7mm}
\hat{H}^{(klm)}_{ij}(\eta) = \sum_{p,q} \hat{E}^{(k)}_{ip} \hat{E}^{(l)}_{pq} \hat{E}^{(m)}_{qj} 
                       h^{(klm)}_{\underline{i} pq \underline{j}}(\eta), \hspace{1cm}
\hat{R}_{ij}(\eta) = \lambda^2 \omega b_5 \left( \hat{V}^{-1} \hat{W}^{(01)} \hat{V} \right)_{ij} 
                                r^{(1)}_{\underline{i} \underline{j}}(\eta).
\eea
The functions $f^{(k)}_{ij}(\eta)$, $g^{(kl)}_{ipj}(\eta)$, $h^{(klm)}_{ipqj}(\eta)$ and
$r^{(k)}_{ij}(\eta)$ are given by
\bea
f^{(k)}_{ij}(\eta) &=& \left\{ \begin{array}{lcl}
 \eta^{a_i} \ln \eta, &\hspace{5mm}& {\rm when}~~ a_j + k - a_i = 0,\\
 \f{1}{a_j + k - a_i} \left( \eta^{a_j + k} - \eta^{a_i} \right) , && {\rm otherwise},
\end{array} \right.\\[2mm]
r^{(k)}_{ij}(\eta) &=& \left\{ \begin{array}{ll}
 \f{1}{2} \eta^{a_i} \ln^2 \eta, & {\rm when}~~ a_j + k - a_i = 0,\\
 \f{1}{a_j + k - a_i} \left( \eta^{a_j+k} \ln \eta - f^{(k)}_{ij}(\eta) \right), & {\rm otherwise},
\end{array} \right.\\[2mm]
g^{(kl)}_{ipj}(\eta) &=& \left\{ \begin{array}{lcl}
r^{(k)}_{ip}(\eta) &\hspace{5mm}& {\rm when}~~ a_j + l - a_p = 0,\\
 \f{1}{a_j + l - a_p} 
 \left( f^{(k+l)}_{ij}(\eta) - f^{(k)}_{ip}(\eta) \right), && {\rm otherwise},
\end{array} \right.\\[2mm]
h^{(klm)}_{ipqj}(\eta) &=& \left\{ \begin{array}{lcl}
\f{1}{6} \eta^{a_i} \ln^3 \eta, 
&& \hspace{-4cm} {\rm when}~~ a_p + k - a_i = a_q + l - a_p = a_j + m - a_q = 0,\\[2mm]
\f{1}{a_p+k-a_i} \left( \f{1}{2} \eta^{a_p+k} \ln^2 \eta - r^{(k)}_{ip}(\eta) \right),
&& \begin{array}{l}  {\rm when}~~ a_p + k - a_i \neq 0 ~~{\rm and} \\[-1mm] 
                     a_q + l - a_p = a_j + m - a_q = 0, \end{array} \\[3mm]
\f{1}{a_q+l-a_p} \left( r^{(k+l)}_{iq}(\eta) - g^{(kl)}_{ipq}(\eta) \right),
&& \begin{array}{l}  {\rm when}~~ a_q + l - a_p \neq 0 ~~{\rm and} \\[-1mm] 
                     \hspace{12mm} a_j + m - a_q = 0, \end{array} \\[2mm]
\f{1}{a_j+m-a_q} \left( g^{(k,l+m)}_{ipj}(\eta) - g^{(kl)}_{ipq}(\eta) \right),
&& {\rm when}~~ a_j + m - a_q \neq 0.
\end{array} \right.
\eea

\subsection{Anomalous dimension matrices \label{sec:ADM}}

In the present Section, we give the ADM's for the four-fermion
operators. When the operators are ordered as in the list $\{ P_1, ...,
P_6, P_9, P_{10}, P_{3Q}, ..., P_{6Q}, P_b \}$, then the matrices that
enter Eq.~(\ref{ADMexp}) have the following generic structure:
\be
\hat{\gamma}^{(nm)} = \left[ \begin{array}{ccccc}
(\hat{\gamma}^{(nm)}_{CC})_{2\times 2} &
(\hat{\gamma}^{(nm)}_{CP})_{2\times 4} &
(\hat{\gamma}^{(nm)}_{CL})_{2\times 2} &
(\hat{\gamma}^{(nm)}_{CQ})_{2\times 4} &
0_{2\times 1}\\[2mm]
0_{4\times 2} &
(\hat{\gamma}^{(nm)}_{PP})_{4\times 4} &
(\hat{\gamma}^{(nm)}_{PL})_{4\times 2} &
(\hat{\gamma}^{(nm)}_{PQ})_{4\times 4} &
0_{4\times 1}\\[2mm]
0_{2\times 2} &
(\hat{\gamma}^{(nm)}_{LP})_{2\times 4} &
(\hat{\gamma}^{(nm)}_{LL})_{2\times 2} &
(\hat{\gamma}^{(nm)}_{LQ})_{2\times 4} &
0_{2\times 1}\\[2mm]
0_{4\times 2} &
(\hat{\gamma}^{(nm)}_{QP})_{4\times 4} &
(\hat{\gamma}^{(nm)}_{QL})_{4\times 2} &
(\hat{\gamma}^{(nm)}_{QQ})_{4\times 4} &
0_{4\times 1}\\[2mm]
0_{1\times 2} &
(\hat{\gamma}^{(nm)}_{BP})_{1\times 4} &
(\hat{\gamma}^{(nm)}_{BL})_{1\times 2} &
(\hat{\gamma}^{(nm)}_{BQ})_{1\times 4} &
(\hat{\gamma}^{(nm)}_{BB})_{1\times 1} 
\end{array} \right].
\ee
However, the pure-QCD ADM's have a much simpler structure
\be
\hat{\gamma}^{(n0)} = \left[ \begin{array}{ccccc}
(\hat{\gamma}^{(n0)}_{CC})_{2\times 2} &
(\hat{\gamma}^{(n0)}_{CP})_{2\times 4} &
0_{2\times 2} &
0_{2\times 4} &
0_{2\times 1}\\[2mm]
0_{4\times 2} &
(\hat{\gamma}^{(n0)}_{PP})_{4\times 4} &
0_{4\times 2} &
0_{4\times 4} &
0_{4\times 1}\\[2mm]
0_{2\times 2} &
0_{2\times 4} &
0_{2\times 2} &
0_{2\times 4} &
0_{2\times 1}\\[2mm]
0_{4\times 2} &
(\hat{\gamma}^{(n0)}_{QP})_{4\times 4} &
0_{1\times 2} &
(\hat{\gamma}^{(n0)}_{QQ})_{4\times 4} &
0_{4\times 1}\\[2mm]
0_{1\times 2} &
(\hat{\gamma}^{(n0)}_{BP})_{1\times 4} &
0_{1\times 2} &
0_{1\times 4} &
(\hat{\gamma}^{(n0)}_{BB})_{1\times 1}
\end{array} \right].
\ee
Moreover, four additional blocks vanish in $\hat{\gamma}^{(01)}$
\be
\hat{\gamma}^{(01)}_{CP} = 0, \hspace{2cm}
\hat{\gamma}^{(01)}_{PP} = 0, \hspace{2cm}
\hat{\gamma}^{(01)}_{LP} = 0, \hspace{2cm}
\hat{\gamma}^{(01)}_{BP} = 0.
\ee
We need to know all the non-vanishing blocks of
$\hat{\gamma}^{(10)}$ and $\hat{\gamma}^{(20)}$:\\[-7mm]
\bea && \hspace{-15mm}
\hat{\gamma}^{(10)}_{CC} =
\left[ \begin{array}{cc} 
-4 & \f{8}{3} \\[1mm]
12 &     0    
\end{array} \right]\!, 
\hspace{11mm}
\hat{\gamma}^{(10)}_{CP} =
\left[ \begin{array}{cccc} 
0     &   -\f{2}{9} &      0    &     0 \\[1mm]
0     &    \f{4}{3} &      0    &     0     
\end{array} \right]\!,
\hspace{11mm}
\hat{\gamma}^{(10)}_{PP} =
\left[ \begin{array}{cccc} 
       0     &  -\f{52}{3} &      0    &     2     \\[1mm]
  -\f{40}{9} & -\f{100}{9} &  \f{4}{9} &  \f{5}{6} \\[1mm]
       0     & -\f{256}{3} &      0    &    20     \\[1mm]
 -\f{256}{9} &   \f{56}{9} & \f{40}{9} & -\f{2}{3} 
\end{array} \right]\!, 
\nnb\\[-4mm] && \hspace{-15mm}
\hat{\gamma}^{(10)}_{QP} =
\left[ \begin{array}{cccc} 
 0&   -\f{8}{9}&         0&         0 \\[1mm]
 0&  \f{16}{27}&         0&         0 \\[1mm]
 0& -\f{128}{9}&         0&         0 \\[1mm]
 0& \f{184}{27}&         0&         0
\end{array} \right]\!,
\hspace{5mm}
\hat{\gamma}^{(10)}_{QQ} =
\left[ \begin{array}{cccc} 
           0&         -20&         0&         2 \\[1mm]
  -\f{40}{9}&  -\f{52}{3}&  \f{4}{9}&  \f{5}{6} \\[1mm]
           0&        -128&         0&        20 \\[1mm]
 -\f{256}{9}& -\f{160}{3}& \f{40}{9}& -\f{2}{3} 
\end{array} \right]\!,
\nnb\\[-18mm] && \hspace{11cm}
\hat{\gamma}^{(10)}_{BB} = [ 4 ], 
\\[3mm] && \hspace{105mm}
\hat{\gamma}^{(10)}_{BP} =
\left[ \begin{array}{cccc} 
0     &    \f{4}{3} &      0    &     0     
\end{array} \right]\!,
\nnb \\ \nnb
\eea
\vspace*{-8mm}
\bea && 
\hat{\gamma}^{(20)}_{CC} =
\left[ \begin{array}{cc} 
-\f{355}{9} & -\f{502}{27} \\[1mm] 
 -\f{35}{3} &   -\f{28}{3} 
\end{array} \right],
\hspace{2cm}
\hat{\gamma}^{(20)}_{CP} =
\left[ \begin{array}{cccc} 
 -\f{1412}{243} &  -\f{1369}{243} &    \f{134}{243} &   -\f{35}{162} \\[1mm]
   -\f{416}{81} &    \f{1280}{81} &      \f{56}{81} &     \f{35}{27} 
\end{array} \right],
\nnb \\[3mm] && \hspace{-15mm}
\hat{\gamma}^{(20)}_{PP} =
\left[ \begin{array}{cccc} 
   -\f{4468}{81} &  -\f{31469}{81} &     \f{400}{81} &  \f{3373}{108} \\[1mm]
  -\f{8158}{243} & -\f{59399}{243} &    \f{269}{486} & \f{12899}{648} \\[1mm]
 -\f{251680}{81} & -\f{128648}{81} &   \f{23836}{81} &   \f{6106}{27} \\[1mm]
  \f{58640}{243} & -\f{26348}{243} & -\f{14324}{243} & -\f{2551}{162} 
\end{array} \right], 
\hspace{5mm}
\hat{\gamma}^{(20)}_{QP} = 
\left[ \begin{array}{cccc} 
 \f{832}{243}   & -\f{4000}{243}  & -\f{112}{243}  & -\f{70}{81}    \\[1mm]
 \f{3376}{729}  &  \f{6344}{729}  & -\f{280}{729}  &  \f{55}{486}   \\[1mm]
 \f{2272}{243}  & -\f{72088}{243} & -\f{688}{243}  & -\f{1240}{81}  \\[1mm]
 \f{45424}{729} &  \f{84236}{729} & -\f{3880}{729} &  \f{1220}{243}  
\end{array} \right], 
\nnb \\[3mm] && \hspace{-15mm}
\hat{\gamma}^{(20)}_{QQ} =
\left[ \begin{array}{cccc} 
 -\f{404}{9}    & -\f{3077}{9}   &  \f{32}{9}    &  \f{1031}{36}  \\[1mm]
 -\f{2698}{81}  & -\f{8035}{27}  & -\f{49}{162}  &  \f{4493}{216} \\[1mm]
 -\f{19072}{9}  & -\f{14096}{9}  &  \f{1708}{9}  &  \f{1622}{9}   \\[1mm]
  \f{32288}{81} & -\f{15976}{27} & -\f{6692}{81} & -\f{2437}{54}  
\end{array} \right], 
\hspace{.5cm}
\begin{array}{c}
\hat{\gamma}^{(20)}_{BP} = 
\left[ \begin{array}{cccc} 
 -\f{1576}{81}    & \f{446}{27}   &  \f{172}{81}    &  \f{40}{27}  
\end{array} \right], \\ \\
\hat{\gamma}^{(20)}_{BB} = 
\left[ \begin{array}{c} 
 \f{325}{9}
\end{array} \right].
\end{array}
\eea 
Almost all the blocks of $\hat{\gamma}^{(01)}$ are necessary: \hspace{2cm}
$\hat{\gamma}^{(01)}_{BL} =
\left[ \begin{array}{cc} 
  \f{16}{9} & 0       
\end{array} \right],$
\bea && \hspace{-15mm}
\hat{\gamma}^{(01)}_{CC} =
\left[ \begin{array}{cc} 
 -\f{8}{3}&         0 \\[1mm]
         0& -\f{8}{3}
\end{array} \right]\!,
\hspace{4mm}
\hat{\gamma}^{(01)}_{CL} =
\left[ \begin{array}{cc} 
 -\f{32}{27}&         0\\[1mm] 
   -\f{8}{9}&         0
\end{array} \right]\!, 
\hspace{4mm}
\hat{\gamma}^{(01)}_{CQ} =
\left[ \begin{array}{cccc} 
   \f{32}{27}&          0&          0&          0 \\[1mm]
     \f{8}{9}&          0&          0&          0
\end{array} \right]\!,
\hspace{4mm}
\hat{\gamma}^{(01)}_{LL} =
\left[ \begin{array}{cc} 
  8 &  -4 \\[1mm]
 -4 &   0
\end{array} \right]\!,
\nnb\\[2mm] && \hspace{-15mm}
\hat{\gamma}^{(01)}_{PQ} =
\left[ \begin{array}{cccc} 
    \f{76}{9}&          0&  -\f{2}{3}&          0 \\[1mm]
  -\f{32}{27}&  \f{20}{3}&          0&  -\f{2}{3} \\[1mm]
   \f{496}{9}&          0& -\f{20}{3}&          0 \\[1mm]
 -\f{512}{27}& \f{128}{3}&          0& -\f{20}{3} \\[1mm]
\end{array} \right]\!,
\hspace{5mm}
\hat{\gamma}^{(01)}_{PL} =
\left[ \begin{array}{cc} 
  -\f{16}{9}&         0\\[1mm] 
  \f{32}{27}&         0\\[1mm] 
 -\f{112}{9}&         0\\[1mm] 
 \f{512}{27}&         0
\end{array} \right]\!, 
\hspace{5mm}
\hat{\gamma}^{(01)}_{QL} =
\left[ \begin{array}{cc} 
  -\f{272}{27} &         0\\[1mm] 
   -\f{32}{81} &         0\\[1mm] 
 -\f{2768}{27} &         0\\[1mm] 
  -\f{512}{81} &         0
\end{array} \right]\!. 
\eea

From $\hat{\gamma}^{(02)}$ we need only the mixing of $P_1$, ..., $P_6$ 
into $P_9$ and $P_{10}$:\\[-3mm]
\bea && 
\hat{\gamma}^{(02)}_{CL} =
\left[ \begin{array}{cc} 
 -\f{11680}{2187} & -\f{416}{81} \\[1mm] 
 -\f{2920}{729}   &  -\f{104}{27}
\end{array} \right],
\hspace{2cm}
\hat{\gamma}^{(02)}_{PL} =
\left[ \begin{array}{cc} 
-\f{39752}{729}  & -\f{136}{27} \\[1mm]
 \f{1024}{2187}  & -\f{448}{81} \\[1mm]
-\f{381344}{729} & -\f{15616}{27} \\[1mm]
 \f{24832}{2187} & -\f{7936}{81}
\end{array} \right].
\eea
The necessary entries of $\hat{\gamma}^{(11)}$ read:
\bea && \hspace{-15mm}
\hat{\gamma}^{(11)}_{CC} =
\left[ \begin{array}{cc} 
\f{169}{9} & \f{100}{27} \\[2mm] 
\f{50}{3}  &  -\f{8}{3}
\end{array} \right]\!,
\hspace{4mm}
\hat{\gamma}^{(11)}_{CP} =
\left[ \begin{array}{cccc} 
0 & \f{254}{729}  & 0 & 0 \\[2mm]
0 & \f{1076}{243} & 0 & 0
\end{array} \right]\!,
\hspace{4mm}
\hat{\gamma}^{(11)}_{CQ} =
\left[ \begin{array}{cccc} 
\f{2272}{729}  & \f{122}{81}  & 0 & \f{49}{81} \\[2mm]
-\f{1952}{243} & -\f{748}{27} & 0 & \f{82}{27}
\end{array} \right]\!,
\nnb\\[3mm] && \hspace{-15mm}
\hat{\gamma}^{(11)}_{PP} =
\left[ \begin{array}{cccc} 
 0           & \f{11116}{243}  &   0          & -\f{14}{3}  \\[2mm]
\f{280}{27}  & \f{18763}{729}  & -\f{28}{27}  & -\f{35}{18} \\[2mm]
 0           & \f{111136}{243} &   0          & -\f{140}{3} \\[2mm]
\f{2944}{27} & \f{193312}{729} & -\f{280}{27} & -\f{175}{9}
\end{array} \right]\!,
\hspace{6mm}
\hat{\gamma}^{(11)}_{PQ} =
\left[ \begin{array}{cccc} 
-\f{23488}{243}  & \f{6280}{27}   &  \f{112}{9}  & -\f{538}{27}   \\[2mm]
 \f{31568}{729}  & \f{9481}{81}   & -\f{92}{27}  & -\f{1012}{81}  \\[2mm]
-\f{233920}{243} & \f{68848}{27}  &  \f{1120}{9} & -\f{5056}{27}  \\[2mm]
 \f{352352}{729} & \f{116680}{81} & -\f{752}{27} & -\f{10147}{81} \\[2mm]
\end{array} \right]\!,
\nnb\\[2mm] && \hspace{-15mm}
\hat{\gamma}^{(11)}_{PL} =
\left[\! \begin{array}{cc} 
  -\f{6752}{243}&  0\\[2mm] 
  -\f{2192}{729}&  0\\[2mm] 
 -\f{84032}{243}&  0\\[2mm] 
 -\f{37856}{729}&  0\\[2mm]
\end{array} \!\right]\!, 
\hspace{2mm}
\hat{\gamma}^{(11)}_{QL} =
\left[\! \begin{array}{cc} 
-\f{24352}{729}    & 0 \\[2mm] 
 \f{54608}{2187}   & 0 \\[2mm] 
-\f{227008}{729}   & 0 \\[2mm] 
 \f{551648}{2187}  & 0
\end{array} \!\right]\!,
\nnb\\[-2.3cm] && \hspace{7cm}
\begin{array}{c}\vspace*{-7pt}
\hat{\gamma}^{(11)}_{CL} = \left[\! \begin{array}{cc}
    -\f{2272}{729}&  0\\[2mm]
    \f{1952}{243}&  0
\end{array} \!\right]\!, 
\hspace{2mm}
\hat{\gamma}^{(11)}_{LL} =
\left[\!\! \begin{array}{cc} 
  0 & 16  \\[1mm]
 16 &  0 
\end{array} \!\right]\!,\\ \\
\hat{\gamma}^{(11)}_{BL} =
\left[ \begin{array}{cc} 
 -\f{8}{9}  &  0
\end{array} \right]\!.
\end{array}
\eea
Finally, the relevant 3-loop anomalous dimensions yield
\cite{Gorbahn:2004my}
\bea
&& \hspace{-15mm}
\hat{\gamma}^{(30)}_{CC} = 
\left[ \begin{array}{cc}
-\f{12773}{18}+\f{1472\zeta(3)}{3}& \f{745}{9}-\f{4288\zeta(3)}{9}\\[1mm] 
\f{1177}{2}-2144\zeta(3)& 306-224\zeta(3)
\end{array} \right], 
\hspace{11mm}
\hat{\gamma}^{(21)}_{CL} =
\left[ \matrix{ -\frac{1359190}{19683}   + \frac{6976\,\zeta(3)}{243} & 0 \cr - \frac{229696}{6561}  - 
   \frac{3584\,\zeta(3)}{81} & 0 \cr  } \right],\\[4mm]
&& \hspace{-15mm}
\hat{\gamma}^{(30)}_{CP} = 
\left[ \begin{array}{cccc}
\f{63187}{13122}& -\f{981796}{6561}& -\f{202663}{52488}& -\f{24973}{69984}\\[1mm] 
\f{110477}{2187}& \f{133529}{8748}& -\f{42929}{8748}& \f{354319}{11664}
\end{array} \right]
+ \zeta(3)
\left[ \begin{array}{cccc}
-\f{1360}{81}& -\f{776}{81}& \f{124}{81}& \f{100}{27}\\[1mm] 
\f{2720}{27}& -\f{2768}{27}& -\f{248}{27}& -\f{110}{9}
\end{array} \right],\\[4mm]
&& \hspace{-15mm}
\hat{\gamma}^{(30)}_{PP} = 
\left[ \begin{array}{cccc}
-\f{3572528}{2187}& -\f{58158773}{8748}& \f{552601}{4374}& \f{6989171}{11664}\\[1mm] 
-\f{1651004}{6561}& -\f{155405353}{52488}& \f{1174159}{52488}& \f{10278809}{34992}\\[1mm] 
-\f{147978032}{2187}& -\f{168491372}{2187}& \f{11213042}{2187}& \f{17850329}{2916}\\[1mm] 
\f{136797922}{6561}& -\f{72614473}{13122}& -\f{9288181}{6561}& -\f{16664027}{17496}
\end{array} \right]\\[4mm]
&& \hspace{-11mm}
+ \zeta(3)
\left[ \begin{array}{cccc}
-\f{608}{27}& \f{61424}{27}& -\f{496}{27}& -\f{2821}{9}\\[1mm] 
 \f{88720}{81}& \f{54272}{81}& -\f{9274}{81}& -\f{3100}{27}\\[1mm] 
 \f{87040}{27}& \f{324416}{27}& -\f{13984}{27}& -\f{31420}{9}\\[1mm] 
 \f{721408}{81}& -\f{166432}{81}& -\f{95032}{81}& -\f{7552}{27}
\end{array} \right], \hspace{11mm}
\hat{\gamma}^{(21)}_{PL} =
\left[ \matrix{ 
 -\frac{1290092}{6561}   + \frac{3200\,\zeta(3)}{81} & 0 \cr - \frac{819971}{19683}  - 
   \frac{19936\,\zeta(3)}{243} & 0 \cr - \frac{16821944}{6561}  + 
   \frac{30464\,\zeta(3)}{81} & 0 \cr - \frac{17787368}{19683}   - 
   \frac{286720\,\zeta(3)}{243} & 0 \cr  } \right].
\eea
The three-loop ADM's have no influence on the logarithmically-enhanced
QED corrections at the considered order but are necessary for the NNLO
QCD corrections. As far as the one- and two-loop ADM's are concerned,
we have calculated all of them, and our results agree with
Ref.~\cite{Bobeth:2003at}.

\subsection{Wilson coefficients at the low scale \label{sec:lowscaleWC}}
From the solution to the RGE in Section~\ref{sec:RGE}, we obtain the Wilson
coefficients at the scale $\mu_b \sim m_b$ as  truncated series  in $\as
(\mu_0)$ and $\kappa (\mu_0)$. We then use Eq.~(\ref{alpharun}) to express the
couplings at the high scale in terms of $\as (\mu_b)$ and $\kappa (\mu_b)$.
 For $\as$, the simple relation 
\be
\displaystyle \as(\mu_0) = \eta \; \as(\mu_b)
\ee
holds to all orders. In order to obtain the running of $\kappa$, we invert
Eq.~(\ref{alpharun}), treating  $v_s$ and $\eta$  as quantities of order
${\cal O}(1)$,  which gives 
\be
\displaystyle \kappa(\mu_0) = \frac{\kappa(\mu_b)}{\eta} +
\frac{\beta^e_{00}}{\beta^s_{00}} \, \frac{1-\eta}{\eta^2} \, \kappa^2(\mu_b)
+\frac{\ln \eta}{\eta}\left[\frac{\beta^e_{00} \, 
\beta^s_{10}}{(\beta^s_{00})^2}-\frac{\beta^e_{01}}{\beta^s_{00}}\right] \, \as (\mu_b)
\, \kappa^2(\mu_b) +  {\cal O} (\kappa^2 \as^2, \kappa^3) \; .
\ee
The final expression for the Wilson coefficients at the low scale is:
\bea
C_k (\mu_b) = \sum_{n,m=0}^2 \as(\mu_b)^n \kappa(\mu_b)^m \; C_k^{(nm)} (\mu_b) 
+ {\cal O}(\as^3,\kappa^3)\; ,
\eea
where $\vec{C}^{(n,m)}$ are functions of only ~ $\eta =
\as(\mu_0)/\as(\mu_b)$,~  $s_W^2$ and ratios of the heavy
masses. At order ${\cal O}(\as^2 \kappa^2)$, we keep only those
contributions to $C_9$ and $C_{10}$ that are proportional to
$m_t^4/(M_W^4 s_W^4)$.

The matching conditions, anomalous dimensions and RG-equations
presented in Sections~\ref{sec:matching}--\ref{sec:ADM} do not include
the two dipole operators $P_{7,8}$. For those two operators, it is
more convenient to consider the so-called effective coefficients
\bea 
C_7^{\rm eff} (\mu_b) & \equiv &  
C_7 (\mu_b) + \sum_{i=3}^6 y_i \left[ C_i (\mu_b) -{1\over 3} C_{iQ}(\mu_b)\right] \nnb\\
&=&
C_7^{(00){\rm eff}} (\mu_b) + \as(\mu_b) \; C_7^{(10){\rm eff}} (\mu_b)  
+ \kappa (\mu_b) \; C_7^{(01){\rm eff}} (\mu_b) 
\nnb\\ &&
+ \as(\mu_b) \kappa (\mu_b) \; C_7^{(11){\rm eff}} (\mu_b) 
+ {\cal O}(\as^2,\kappa^2)\;, \label{c7eff}\\
C_8^{\rm eff} (\mu_b) & \equiv &
C_8 (\mu_b) + \sum_{i=3}^6 z_i \left[ C_i (\mu_b) -{1\over 3} C_{iQ}(\mu_b)\right]
= 
C_8^{(00){\rm eff}} (\mu_b)+ {\cal O}(\as,\kappa) \label{c8eff}
\eea
where, in dimensional regularization with fully anticommuting
$\gamma_5$,
$y=(0,0,-{1\over 3},-{4\over 9},-{20\over 3},-{80\over9})$
and 
$z=(0,0,1,-{1\over 6},20,-{10\over 3})$.
The effective coefficients $C_7^{(00){\rm eff}}(\mu_b)$,
$C_7^{(10){\rm eff}}(\mu_b)$ and $C_8^{(00){\rm eff}} (\mu_b)$ can be
found in Eqs.~(10)--(22) of Ref.~\cite{Chetyrkin:1996vx}, while
$C_7^{(01){\rm eff}} (\mu_b)$ can be found in Eq.~(12) of
Ref.~\cite{Baranowski:1999tq}.

Following Ref.~\cite{Bobeth:2003at}, we take into account the  {\em complete} 
${\cal O}(\as\kappa)$ term in $C_7^{\rm eff}(\mu_b)$  rather than only its
  $m_t^2/(M_W^2s^2_W)$-enhanced part (as Section~\ref{sec:branchingratio}
  would imply).  An explicit expression for $C_7^{(11){\rm eff}}(\mu_b)$ can
be found in Eq.~(30) of Ref.~\cite{Gambino:2001au}.
\begin{table}[t]
\begin{center}
\small 
\begin{tabular}{|c|c|c|c|}\hline
      & (00)  & (01) & (10)  \\ \hline
$C_1^{(nm)}$ 
      & [ -0.763 , -0.544 , -0.379 ]
      & [-0.180, -0.0835, -0.0378]
      & [ 13.764 , 14.943 , 16.066 ]  \\ 
$C_2^{(nm)}$ 
      & [ 1.054 , 1.029 , 1.015 ]    
      & [0.248, 0.158, 0.101] 
      & [ -1.746 , -1.376 , -1.050 ]  \\
$C_3^{(nm)}$ 
      & [ -1.10 , -0.571 , -0.283 ]$ 10^{-2}$ 
      & [-1.22, -0.400, -0.125]$ 10^{-3}$ 
      & [ 5.28  , 7.98  , 8.38 ]$ 10^{-2}$  \\ 
$C_4^{(nm)}$ 
      & [ -0.113 , -0.0741 , -0.0486 ]
      & [-1.62, -0.697, -0.297]$ 10^{-2}$ 
      & [ -0.690 , -0.343 , -0.143 ] \\ 
$C_5^{(nm)}$ 
      & [ 1.04  , 0.547 ,  0.274 ]$ 10^{-3}$ 
      & [1.17, 0.387, 0.122]$ 10^{-4}$ 
      & [ -1.60 , -1.55 , -1.36 ]$ 10^{-2}$\\ 
$C_6^{(nm)}$ 
      & [ 2.32 , 1.17 , 0.563 ]$ 10^{-3}$ 
      & [2.51, 0.801, 0.245]$ 10^{-4}$ 
      & [ -0.656 , -1.92 , -2.17 ]$ 10^{-2}$ \\ \hline
$C_{3Q}^{(nm)}$ 
         &   0  
         & [-5.03, -3.72, -2.66]$ 10^{-2}$ 
         &  0  \\ 
$C_{4Q}^{(nm)}$ 
         &  0  
         & [-2.13, -1.04, -0.49]$ 10^{-2}$ 
         &  0  \\ 
$C_{5Q}^{(nm)}$ 
         &  0 
         & [-6.08, -1.71, -0.43]$ 10^{-6}$ 
         & 0  \\ 
$C_{6Q}^{(nm)}$ 
         &  0 
         & [2.12, 1.03, 0.485]$ 10^{-3}$ 
         &  0  \\ \hline
$C_b^{(nm)}$    
         &   0  
         &   0  
         &   0   \\ \hline
\end{tabular}
\caption{Numerical values of the relevant $C_k^{(nm)}(\mu_b)$ ($k\neq 7,8,9,10$) for 
$\mu_b=[ 2.5 , 5 , 10 ]\,\gev$.  \label{tab:wcmub1}}
\end{center}
\end{table}
\begin{table}[ht]
\begin{center}
\small
\begin{tabular}{|c|c|c|}\hline
 & $C_7^{(nm){\rm eff}}(\mu_b)$ & $C_8^{(nm){\rm eff}}(\mu_b)$ \\ \hline
(00) & [ -0.362 , -0.320 , -0.285 ]   & [ -0.168 , -0.151 , -0.138 ]  \\ 
(01) & [ 3.20 , 3.33 , 2.82 ]$10^{-2}$  & $-$  \\
(10) & [ 1.625 , 1.171 , 0.690 ]   & $-$  \\ 
(11) & [ 4.132 , 4.314 , 4.397 ]  & $-$  \\ 
\hline
\end{tabular}
\caption{Numerical values of the relevant $C_{7,8}^{(nm){\rm eff}}(\mu_b)$
for $\mu_b=[ 2.5 , 5 , 10 ]\,\gev$.  \label{tab:wcmub2}}
\end{center}
\end{table}
\begin{table}[ht]
\begin{center}
\small
\begin{tabular}{|c|c|c|}\hline
 & $C_9^{(nm)}(\mu_b)$ & $C_{10}^{(nm)}(\mu_b)$ \\ \hline
(00) & 0  & 0  \\ 
(01) & [ 5.025 , 3.722 , 2.664 ]$ 10^{-2}$  & 0  \\
(10) &  0  & 0  \\ 
(11) & [ 2.003 , 1.934 , 1.863 ]   & [ -4.222 , -4.222 , -4.222 ]  \\ 
(20) &  0  & 0  \\ 
(02) & [ 0.376 , 0.208 , 0.104 ]$ 10^{-2}$  & [ 1.081 , 0.489 , 0.218 ]$ 10^{-2}$   \\
(12) & [ -6.614 , -4.317 , -2.810 ]   & [ -5.854 , -3.798 , -2.458]  \\ 
(21) & [ 5.645 , 3.538 , 1.193 ]   & [ 5.105 , 6.380 , 7.631]  \\
(22) & [ 36.814 , 27.320 , 20.275 ]   & [ -32.014 , -36.090 , -39.764]  \\
\hline
\end{tabular}
\caption{Numerical values of the relevant $C_{9,10}^{(nm)}(\mu_b)$
for $\mu_b=[ 2.5 , 5 , 10 ]\,\gev$.  \label{tab:wcmub3}}
\end{center}
\end{table}
In Tables~\ref{tab:wcmub1} -- \ref{tab:wcmub3}, we present the
relevant $C_k^{(nm)}(\mu_b)$. We fix the input parameters to their
central values (specified in sec.~\ref{sec:branchingratio}) and choose
$\mu_b=[2.5,5,10]\;\gev$ and $\mu_0=120\;\gev$.

\section{Matrix elements I}
\label{sec:melem}
Once $\vec{C}^{(n,m)}(\mu_b)$ is found, one needs to calculate the on-shell
$b \to s l^+ l^-$ matrix elements $\me{P_i}$ of the corresponding operators.
In the present section, we consider those parts of the matrix elements that
originate from diagrams with no photons inside loops and/or bremsstrahlung
photons. These parts are unrelated to the $\ln (m_b^2/m_\ell^2)$-enhanced
correction to the decay width.

One-loop penguin contractions of the 4-fermion operators give the following
contributions to the matrix elements:
\bea
\me{P_i}^{\rm peng} & = &  M_i^9  \me{P_9}_{\rm tree} +
                M_i^7  \frac{\me{P_7}_{\rm tree}}{\as(\mu_b) \kappa (\mu_b)} + 
                M_i^{10}  \me{P_{10}}_{\rm tree} \;. 
\label{matelPi}
\eea
The above formula holds also for the tree-level matrix element of $P_7$, the
one-loop matrix element of $P_8$, and for those parts of the two-loop ${\cal
  O}(\alpha_s\alpha_{\rm em})$ matrix elements of the 4-quark operators where
the gluon couples to the closed quark loop.  The coefficients
$ M_i^A$ are summarized in Table~\ref{tab:hiA} 
\begin{table}
\begin{center}
\begin{tabular}{|l|c|c|c|} \hline
 & $ M_i^9$ &  $ M_i^7$ &  $ M_i^{10}$ \\ \hline 
i=1,2 & 
$\as \kappa \; f_{i} (\hat s) - \as^2 \kappa \; F_i^9 (\hat s) $ &
$- \as^2 \kappa \; F_i^7 (\hat s) $ &
0 \\       
i=3-6,3Q-6Q,b &
$\as \kappa \; f_{i} (\hat s) $ & 
0 & 0 \\
i=7 &
0&
$\as \kappa$ &0\\
i=8 &
$- \as^2 \kappa \; F_8^9 (\hat s) $ &
$- \as^2 \kappa \; F_8^7 (\hat s) $ & 0 \\ 
i=9 &
$1+\as \kappa \; f_9^{\rm pen} (\hat s) $ &
0 & 0\\
i=10 &
0 & 0 & 1 \\\hline
\end{tabular}
\caption{Coefficients $ M_i^A$ that appear in Eq.~(\ref{matelPi}). 
Here, $\as$ and $\kappa$ are taken at the scale $\mu_b$. \label{tab:hiA}}
\end{center}
\end{table}
 in terms of the functions $F_i^A(\hat s)$ and  
\bea 
f_i (\hat s) &=& 
\gamma^{(01)}_{i9} \ln \f{m_b}{\mu_b}
+ \rho_i^c \left( g(y_c) + \f{8}{9} \ln \f{m_b}{m_c} \right)
+ \rho_i^b g(y_b) + \rho_i^0 \left( \ln \s \!-\! i\pi \right) + \rho_i^\# \; ,
\label{4melem}\\
f_9^{\rm pen} (\hat s) & = &
8 \ln \f{m_b}{\mu_b} - 3 \left( g(y_\tau) + \f{8}{9} \ln \f{m_b}{m_\tau} \right) 
       + \f{8}{3} \left( \ln \s - i\pi \right) - \f{40}{9} \; .
\eea
Here,  $y_a = 4 m_{a,{\rm pole}}^2/m_{\ell^+ \ell^-}^2$,  the function
$g(y)$ is given in Appendix~\ref{sec:loopfunctions}, and the numbers $\rho$
are collected in Table~\ref{tab:rhos}.  The functions $F_i^A(\hat s)$ can
  be found in Eqs.~(54)--(56), (71) and (72) of Ref.~\cite{Asatryan:2001zw}
  where they are given in terms of an expansion in $\hat s$ up to ${\cal
    O}(\hat{s}^3)$. In the range of $\hat s$ that we consider here, the
  accuracy of these expansions is excellent, as can be seen in Fig.~8 of
  Ref.~\cite{Ghinculov:2003qd} where the same functions are numerically
  evaluated for arbitrary $\hat s$. 

For what concerns the remaining contributions to the NLO and NNLO QCD matrix
elements of $P_{7,9,10}$, the virtual and real corrections can be effectively
taken into account via the following redefinitions of the squared tree-level
matrix elements in the expression for the decay width:
\bea
\left| \me{P_9}_{\rm tree} \right|^2 
& \Longrightarrow &
\left| \me{P_9}_{\rm tree} \right|^2 
\left[ 1 + 8 \; \as \; \omega_{ 99 }^{(1)} (\hat s) + 16 \;  \as^2 \;
  \omega_{ 99 }^{(2)} (\hat s)
\right] \;, \label{omega9sub}\\
\left| \me{P_{10}}_{\rm tree} \right|^2 
& \Longrightarrow &
\left| \me{P_{10}}_{\rm tree} \right|^2 
\left[ 1 + 8  \; \as \; \omega_{ 1010 }^{(1)} (\hat s) \right]  \;,\\
\left| \me{P_{7}}_{\rm tree} \right|^2 
& \Longrightarrow &
\left| \me{P_{7}}_{\rm tree} \right|^2 
\left[ 1 + 8  \; \as \; \omega_{ 77 }^{(1)}(\hat s) \right]  \;,\\
{\rm Re} \left(\me{P_{7}}_{\rm tree} \me{P_9}_{\rm tree}^*\right)
& \Longrightarrow &
{\rm Re} \left(\me{P_{7}}_{\rm tree} \me{P_9}_{\rm tree}^* \right)
\left[ 1 + 8  \; \as \; \omega_{79}^{(1)}(\hat s) \right] \;,
\label{o79mod}
\eea
where the functions $\omega_{ij}^{(n)} (\hat s)$ calculated in
Refs.~\cite{Asatryan:2001zw,Bobeth:2003at} are listed in
Appendix~\ref{sec:loopfunctions}.

The remaining contributions to the NNLO matrix elements of the 4-quark
operators originate from diagrams where the gluon does not couple to the quark
loop. Thus,  they are  given by the same functions of $\hat s$ as in
Eq.~(\ref{omega9sub}).
\begin{table}[t]
\begin{displaymath}
\hspace*{15mm} 
\begin{array}{|c|rrrrrrrrrrr|}
\hline
& P_1 & P_2 & P_3 & P_4 & P_5 & P_6 & P_3^Q & P_4^Q & P_5^Q & P_6^Q &
P^b \\
\hline
&&&&&&&&&&& \\[-3mm]
\rho^c & \f{4}{3} & 1 & 6 & 0 & 60 & 0 & 4 & 0 & 40 & 0 & 0 \\[2mm]
\rho^b & 0 & 0 & -\f{7}{2} & -\f{2}{3} & -38 & -\f{32}{3} & \f{7}{6} 
       & \f{2}{9} & \f{38}{3} & \f{32}{9} & -2\\[2mm]
\rho^0  & 0 & 0 & \f{2}{9} & \f{8}{27} & \f{32}{9} & \f{128}{27} & -\f{74}{27} 
        & -\f{8}{81} & -\f{752}{27} & -\f{128}{81} & 0\\[2mm]  
\rho^\# & -\f{16}{27} & -\f{4}{9} & \f{2}{27} & \f{8}{81} & -\f{136}{27} & \f{320}{81} 
        & \f{358}{81} & -\f{8}{243} & \f{1144}{81} & -\f{320}{243} & \f{26}{27} \\[1mm]
\hline
\end{array}
\end{displaymath}
\vspace*{-8mm}
\begin{center}
\caption{ Numbers that occur in the four-quark operator 
          matrix elements in Eq.~(\ref{4melem}). \label{tab:rhos}}
\end{center}
\vspace*{-12mm}
\end{table}

\section{Matrix elements II \label{sec:emcorrp9}} 
In this Section, we calculate those electromagnetic corrections to the matrix
elements of the 4-fermion operators that are responsible for the $\ln
(m_b^2/m_\ell^2)$-enhanced correction to the decay width. In
section~\ref{sec:emcorrP9}, we cover in great detail the calculation of QED
corrections to $\langle P_9 \rangle$. In Section~\ref{sec:otheremcorr}, we give
the logarithmically enhanced QED corrections to the matrix elements of $P_i$
with $i \neq 9$.
\subsection{Corrections to $\me{P_9}$}\label{sec:emcorrP9}
Electromagnetic corrections to the matrix element of $P_9$ are infrared
divergent and must be considered together with the corresponding
bremsstrahlung. The dilepton invariant mass differential decay width is not an
infrared safe object with respect to emission of collinear photons. Hence,
electromagnetic corrections contain an explicit collinear logarithm $\ln
(m_b^2/m_\ell^2)$. The coefficient of this logarithm vanishes when integrated
over the whole phase space but not if we restrict it to the low-$\s$ region.

In this calculation, we adopt the NDR scheme with $D=4-2\epsilon$. The NDR
scheme is suitable for our calculation since no Levi-Civita tensor survives in
divergent terms proportional to $1/\epsilon$ or $1/\epsilon^2$. Thus, all the
Levi-Civita tensors can be evaluated in $D = 4$ and are therefore
well-defined.

In the first step, all the external particles are taken to be on-shell, and,
in addition, all the final state particles are treated as massless ($m_s =
m_\ell = 0$). This implies that all the collinear divergences are 
  dimensionally regularized,  and that the collinear logarithm appears as a
residual $1/\epsilon$. Later, we will be able to re-express such a residue in
terms of $\ln (m_b^2/m_\ell^2)$ using the photonic splitting function of the
 lepton. 

In the next two subsections, we present the calculation of virtual and
bremsstrahlung corrections. In the last one, we show how to change the
collinear regulator from dimensional to the physical mass regularization.

The calculation involves the following kinematical invariants:
$\s_{ij} = 2 \, \frac{p_i \cdot p_j}{m_b^2}$ where $i,j \in
\{l^+,l^-,s,b,\gamma\} \equiv \{1,2,s,b,q\}$.

\begin{figure}
\begin{center}
  \begin{fmffile}{qedvirtual}  
     \hspace*{20pt}\parbox{60mm}{\begin{fmfgraph*}(130,120)
     \fmfleft{b,lminus}
     \fmfright{s,lplus}
     \fmf{fermion,tension=3}{lplus,g2}
     \fmf{fermion,tension=6}{gv1,lminus}
     \fmf{plain,tension=6}{g1,gv1}
     \fmf{phantom,tension=17}{g1,vul}
     \fmf{phantom,tension=17}{g2,vur}
     \fmf{phantom,tension=2260}{vul,vur}
     \fmf{phantom,tension=2260}{g1,g2}
     \fmf{plain,tension=6}{gv2,vul}
     \fmf{fermion,tension=6}{b,gv2}
     \fmf{fermion,tension=3}{vur,s}
     \fmffreeze
     \fmf{photon,label={$\gamma$},label.side=left,label.dist=5}{gv2,gv1}
     \fmffreeze
     \fmf{phantom,label={$P_9$},label.side=right,label.dist=6}{vur,g2}
     \fmffreeze
     \fmfv{label= {$b$},label.dist=4}{b}
     \fmfv{label= {$s$},label.dist=4}{s}
     \fmfv{label= {$l^{+}$},label.dist=4}{lplus}
     \fmfv{label= {$l^{-}$},label.dist=1}{lminus}     
     \fmfv{decor.shape=circle,decor.filled=30,decor.size=.06w}{g2}
     \fmfv{decor.shape=circle,decor.filled=30,decor.size=.06w}{vul}
     \fmffreeze
     \fmfv{decor.shape=cross,decor.size=.06w}{g1}
     \fmfv{decor.shape=cross,decor.size=.06w}{vur}
     \fmffreeze          
     \end{fmfgraph*}}
  \end{fmffile}
  \begin{fmffile}{qedreal}  
\hspace*{20pt}\parbox{60mm}{\begin{fmfgraph*}(130,120)
     \fmfleft{b,lminus}
     \fmfright{s,lplus}
     \fmftop{quant1,quant2,quant3,quant4}
     \fmf{fermion,tension=3}{lplus,g2}
     \fmf{fermion,tension=6}{gbrems,lminus}
     \fmf{plain,tension=6}{g1,gbrems}
     \fmf{phantom,tension=17}{g1,vul}
     \fmf{phantom,tension=17}{g2,vur}
     \fmf{phantom,tension=2260}{vul,vur}
     \fmf{phantom,tension=2260}{g1,g2}
     \fmf{fermion,tension=3}{b,vul}
     \fmf{fermion,tension=3}{vur,s}
     \fmffreeze
     \fmf{photon,label={$\gamma$},label.side=right,label.dist=3}{gbrems,quant2}
     \fmffreeze
     \fmf{phantom,label={$P_9$},label.side=right,label.dist=6}{vur,g2}
     \fmffreeze
     \fmfv{label={$b$},label.dist=2}{b}
     \fmfv{label={$s$},label.dist=2}{s}
     \fmfv{label={$l^{+}$},label.dist=2}{lplus}
     \fmfv{label={$l^{-}$},label.dist=-2}{lminus}
     \fmfv{decor.shape=circle,decor.filled=30,decor.size=.06w}{g2}
     \fmfv{decor.shape=circle,decor.filled=30,decor.size=.06w}{vul}
     \fmffreeze
     \fmfv{decor.shape=cross,decor.size=.06w}{g1}
     \fmfv{decor.shape=cross,decor.size=.06w}{vur}
     \fmffreeze          
     \end{fmfgraph*}}
  \end{fmffile}
\vspace*{10pt}
\caption{Examples of diagrams contributing to the virtual (left) and
real (right) electromagnetic corrections to the matrix element of
$P_9$. \label{fig:virtualplusreal}}
\end{center}
\end{figure}
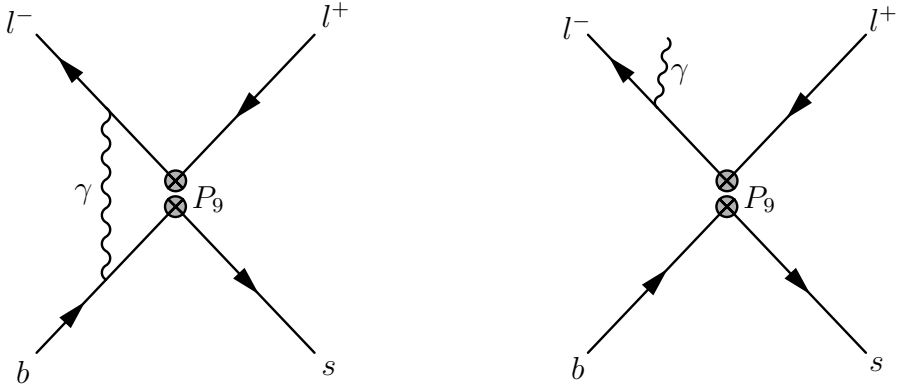

\subsubsection{Virtual corrections}

In order to obtain the virtual corrections, one has to consider one-loop
diagrams of the current-current type. There are in total six such diagrams,
one of which is shown on the left in Figure~\ref{fig:virtualplusreal}. The sum
of the six amplitudes contains infrared and ultraviolet divergences. The
latter cancel after the addition of counterterms. The next step is then to
compute its interference with $\me{P_9}_{\rm tree}$ which yields an expression
$K_V(\sot,\s_{1s},\s_{2s})$. Finally, one has to integrate $K_V$ over the
phase space.  The phase space measure for a three particle massless final
state in $D$ dimensions is given explicitly in~\cite{Gehrmann:2003}. Since
$K_V$ does not depend on angular variables we can immediately integrate them
out
\bea
\widetilde{dPS_3} 
&\equiv&  
\tilde\mu^{4\epsilon} \int_{\Omega} dPS_3 
=
\tilde\mu^{4\epsilon} M_3 (\sot,\s_{1s},\s_{2s}) \, d\sot \, d\s_{1s} \, d\s_{2s} \nnb \\
& = &
\tilde\mu^{4\epsilon} \frac{2^{-8+6\epsilon}  \pi^{-\frac{5}{2}+2\epsilon} 
(m_b^2)^{1-2\epsilon}}{\Gamma(\frac{3}{2}-\epsilon) \, \Gamma(1-\epsilon)} 
(\sot \, \s_{1s} \,\s_{2s})^{-\epsilon} \delta(1-\sot-\s_{1s}-\s_{2s})  
\, d\sot \, d\s_{1s} \, d\s_{2s} \; ,
\eea
where $\tilde\mu^2 = \mu^2 \cdot \exp[\gamma_E-\ln(4\pi)]$. By means of this expression we obtain the final contribution from
virtual corrections via
\bea
T_V 
& \equiv & 
\int \f{\widetilde{dPS_3}}{d\hat{s}_{12}} \; K_V(\sot,\s_{1s},\s_{2s})  \;.
\eea

\subsubsection{Real corrections}
In order to cancel the infrared singularities present in $T_V$ one has
to add the corresponding bremsstrahlung contribution. There are four
diagrams, one of which is shown on the right in
Figure~\ref{fig:virtualplusreal}. Contrary
to the case of gluon bremsstrahlung, the photon couples to all
external legs, which makes the calculation more involved. The sum of
the four amplitudes has to be squared, yielding an expression
$K_R(\sot,\s_{1s},\s_{2s},\s_{1q},\s_{2q},\s_{sq},\s_{tri})$, where
\be 
\s_{tri} \equiv 1 - \sot - \s_{1s} - \s_{2s} = \s_{1q} + \s_{2q} +
\s_{sq} 
\ee 
is the triple invariant. The corresponding phase space measure for the
four particle final state can also be found in~\cite{Gehrmann:2003}.
After integration over angular variables, it reads
\bea
\widetilde{dPS_4}  
\equiv  
\tilde\mu^{6\epsilon} \int_{\Omega} dPS_4  
& = & 
\tilde\mu^{6\epsilon} \cdot \frac{2^{-12+10\epsilon} \,
\pi^{-5+3\epsilon} \,(m_b^2)^{2-3\epsilon}}{\Gamma(\frac{3}{2}-\epsilon) \, 
\Gamma(1-\epsilon) \Gamma(\frac{1}{2}-\epsilon)} \;
d\sot \, d\s_{1s} \, d\s_{1q} \, d\s_{2s} \, d\s_{2q} \, d\s_{sq}  \nnb \\ 
&& \hskip -1cm 
\times \left(-\Delta_4\right)^{-\frac{1}{2}-\epsilon} \cdot \Theta(-\Delta_4)\cdot
\delta(1-\sot-\s_{1s}-\s_{1q}-\s_{2s}-\s_{2q}-\s_{sq}) \, . 
\eea

In the above equation, the Gram determinant is given by
\be
\Delta_4 = (\sot \s_{sq})^2 + (\s_{1s} \s_{2q})^2 +
(\s_{1q}\s_{2s})^2 - 2 \, (\sot \s_{1s} \s_{2q} \s_{sq} + \s_{1s}
\s_{1q}
\s_{2s} \s_{2q} + \sot \s_{1q} \s_{2s} \s_{sq}).
\ee

The phase space measure is completely symmetric in $\{1,2,s,q\}$, but
since we stay differential in $\sot$ we can only make use of the
symmetries $1 \leftrightarrow 2$ and $s \leftrightarrow
q$.\footnote{
  In the terms containing $\s_{tri}$ in the denominator, only the
  $1\leftrightarrow 2$ symmetry can be used.}
The use of these symmetries is, however, essential since the number of
distinct terms in $K_R$ gets reduced significantly. In addition, all terms of
the form $A/(\s_{1q}\s_{sq})$ and $A/(\s_{2q}\s_{sq})$ as well as
$B/(\s_{1q}\s_{tri})$ and $B/(\s_{2q}\s_{tri})$ drop out by means of the $1
\leftrightarrow 2$ symmetry.

Another crucial point is to choose for each term in $\displaystyle
K_R$ the order of integration in a suitable way in order to ensure
that all terms up to and including order $\epsilon^0$ can be found
analytically. Appendix~\ref{app:details} is devoted to this rather
technical issue. The QED bremsstrahlung contribution finally reads
\bea
T_R 
& \equiv & 
\int \f{\widetilde{dPS_4}}{d\hat{s}_{12}} \; K_R(\sot,\s_{1s},\s_{2s},\s_{1q},\s_{2q},\s_{sq},\s_{tri})  \;.
\eea
In the sum of $T_V$ and $T_R$ the $1/\epsilon^2$ terms cancel as well
as the $Q_d^2$ part of the $1/\epsilon$ terms, whereas the collinear
divergences  proportional to  $Q_l^2/\epsilon$ remain.

\subsubsection{From NDR to mass regularization}
\label{NDRtomass}
As we have stated earlier, the differential decay width is not an
infrared safe object with respect to emission of collinear
photons. This means that, as long as the  lepton  is treated as
massless, the sum of virtual and real corrections is not free of
collinear divergences. If we had kept the  lepton  mass different from
zero during the whole calculation, the sum of virtual and real
corrections would have been finite. However, the computation of the
diagrams and the massive phase space integrals in $T_V$ and $T_R$
would have been much more tedious.

The translation of the $1/\epsilon$ pole into a $\ln (m_b^2/m_\ell^2)$
corresponds to changing the regularization scheme and is complicated by the
presence of soft infrared singularities. The correct procedure is to start
with constructing an observable that is infrared safe and, consequently,
regularization scheme independent. Only at this point we can switch to the
$m_\ell$ regulator and obtain our final result. As an intermediate step, we
construct a differential branching ratio where $\hat s$ is identified as
follows:
\bea
\hat s & = & \cases{(p_{\ell_1}+p_{\ell_2}+p_\gamma)^2/m_b^2 & 
        if $\vec p_\gamma$ $||$ ($\vec p_{\ell_1}$ or $\vec p_{\ell_2}$) \cr 
       (p_{\ell_1}+p_{\ell_2})^2/m_b^2 & otherwise.\cr }
\eea
In order to switch to this intermediate observable we must subtract
the {\em collinear} decay width differential in the dilepton invariant
mass and add the same quantity but remaining differential in the
triple invariant.
 
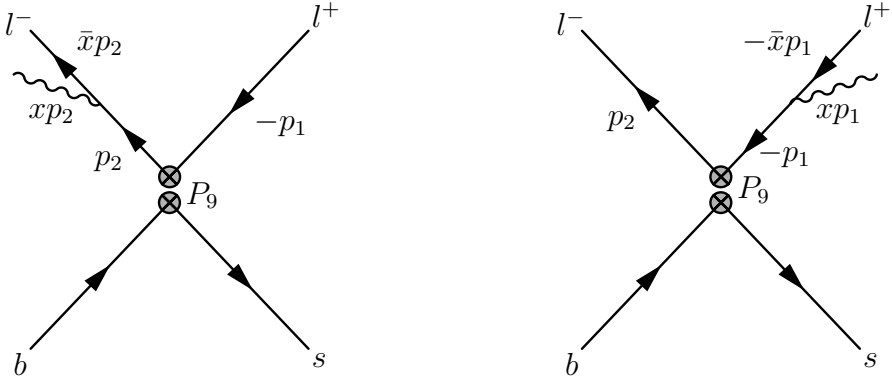
\begin{figure}
\begin{center}
%
 \begin{fmffile}{qedsplitting1}  
     \hspace*{20pt}\parbox{60mm}{\begin{fmfgraph*}(130,120)
     \fmfleft{b,quant1,quant2,quant3,quant4,quant5,quant6,lminus}
     \fmfright{s,lplus}
     \fmf{fermion,label={$-p_1$},label.side=left,tension=3}{lplus,g2}
     \fmf{fermion,label={$\bar x p_2$},label.side=right,tension=6}{gbrems,lminus}
     \fmf{fermion,label={$p_2$},label.side=left,tension=6}{g1,gbrems}
     \fmf{phantom,tension=17}{g1,vul}
     \fmf{phantom,tension=17}{g2,vur}
     \fmf{phantom,tension=2260}{vul,vur}
     \fmf{phantom,tension=2260}{g1,g2}
     \fmf{fermion,tension=3}{b,vul}
     \fmf{fermion,tension=3}{vur,s}
     \fmffreeze
     \fmf{photon,label={$x p_2$},label.side=left}{gbrems,quant6}
     \fmffreeze
     \fmf{phantom,label={$P_9$},label.side=right,label.dist=6}{vur,g2}
     \fmffreeze
     \fmfv{label={$b$},label.dist=2}{b}
     \fmfv{label={$s$},label.dist=2}{s}
     \fmfv{label={$l^{+}$},label.dist=2}{lplus}
     \fmfv{label={$l^{-}$},label.dist=-2}{lminus}
     \fmfv{decor.shape=circle,decor.filled=30,decor.size=.06w}{g2}
     \fmfv{decor.shape=circle,decor.filled=30,decor.size=.06w}{vul}
     \fmffreeze
     \fmfv{decor.shape=cross,decor.size=.06w}{g1}
     \fmfv{decor.shape=cross,decor.size=.06w}{vur}
     \fmffreeze          
     \end{fmfgraph*}}
  \end{fmffile}
 \begin{fmffile}{qedsplitting2}  
     \hspace*{20pt}\parbox{60mm}{\begin{fmfgraph*}(130,120)
     \fmfleft{b,lminus}
     \fmfright{s,quant1,quant2,quant3,quant4,quant5,quant6,lplus}
     \fmf{fermion,label={$p_2$},label.side=left,tension=3}{g1,lminus}
     \fmf{fermion,label={$-\bar x p_1$},label.side=right,tension=6}{lplus,gbrems}
     \fmf{fermion,label={$-p_1$},label.side=left,label.dist=1,tension=6}{gbrems,g2}
     \fmf{phantom,tension=17}{g1,vul}
     \fmf{phantom,tension=17}{g2,vur}
     \fmf{phantom,tension=2260}{vul,vur}
     \fmf{phantom,tension=2260}{g1,g2}
     \fmf{fermion,tension=3}{b,vul}
     \fmf{fermion,tension=3}{vur,s}
     \fmffreeze
     \fmf{photon,label={$x p_1$},label.side=right}{gbrems,quant6}
     \fmffreeze
     \fmf{phantom,label={$P_9$},label.side=right,label.dist=6}{vur,g2}
     \fmffreeze
     \fmfv{label={$b$},label.dist=2}{b}
     \fmfv{label={$s$},label.dist=2}{s}
     \fmfv{label={$l^{+}$},label.dist=2}{lplus}
     \fmfv{label={$l^{-}$},label.dist=-2}{lminus}
     \fmfv{decor.shape=circle,decor.filled=30,decor.size=.06w}{g1}
     \fmfv{decor.shape=circle,decor.filled=30,decor.size=.06w}{vul}
     \fmffreeze
     \fmfv{decor.shape=cross,decor.size=.06w}{g2}
     \fmffreeze
     \fmfv{decor.shape=cross,decor.size=.06w}{vur}
     \fmffreeze          
     \end{fmfgraph*}}
  \end{fmffile}
\vspace*{10pt}
\caption{Splitting function kinematics. The photon is emitted by a
quasi-real lepton. \label{fig:splitting}}
\end{center}
\end{figure}

The calculation of the differential branching ratio in the collinear limit is
done with the help of the NDR-scheme splitting function for the massless 
  lepton. The splitting function in this scheme can be derived from
Refs.~\cite{Terazawa:1973tb,Mele:1990cw} and reads\footnote{We would like to thank Ulrich Haisch, Thomas Gehrmann, and Micha{\l}
Poradzinski for useful discussions on this point.}
\bea
f^{(\epsilon)}_\gamma (x,E) &=&
4 \tilde \alpha_e \left[
\frac{1+(1-x)^2}{x} 
\left( -\frac{1}{2\epsilon} + \ln\frac{E}{\mu}+\ln (2-2x) \right)
+\frac{x}{2}-\frac{(2-x)^2}{2 x} \ln \frac{2-x}{x} \right] \;, \nnb\\
\eea
where $E$ is the energy of the incoming  lepton  and $x E$ is the energy
of the emitted photon. See Fig.~\ref{fig:splitting} for a pictorial view of the kinematics.

The fully differential decay width in the collinear limit is given by (here
and in the following we omit the factor $8 G_F^2 |V_{tb} V_{ts}|^2$ stemming
from the effective Lagrangian):
\bea
{\rm d} \Gamma_{\rm coll}^{(\epsilon)} (\sot,\s_{1s},\s_{2s},x)
&=&
(2 m_b)^{-1} \,
\left[ f^{(\epsilon)}_\gamma (x,E_1) + f^{(\epsilon)}_\gamma (x,E_2) \right] \,
\left| \me{P_9}_{\rm tree} \right|^2 \, dPS_3 \, dx\nnb \\
& = & 
m_b^{-1} \, f^{(\epsilon)}_\gamma (x,E_1) \, \left| \me{P_9}_{\rm tree} \right|^2 \, dPS_3  \, dx\; ,
\label{fullydifferentialwidth}
\eea
where $x,\sot,\s_{1s},\s_{2s} \in [0,1]$, $E_1 = m_b (1-\s_{2s})/2$
and we used the $\ell_1 \leftrightarrow \ell_2$ symmetry. The
collinear decay width differential in the triple invariant 
($\s = (p_{\ell_1}+p_{\ell_2}+p_\gamma)^2/m_b^2$) reads
\bea
\frac{{\rm d} \Gamma^{(\epsilon)}_{\rm coll,3}}{d\s}
=
m_b^{-1}  \, \int_0^1 dx \, \int_0^1 d\s_{1s} \, \int_0^1 d\s_{2s} \, M_3(\s,\s_{1s},\s_{2s})
 \; f^{(\epsilon)}_\gamma (x,E_1) \; \left| \me{P_9}_{\rm tree} \right|^2_{\sot \rightarrow \s} \; .
\eea
The collinear decay width differential in the dilepton invariant mass ($\s =
(p_1+\bar x p_2)^2/m_b^2$) reads
\bea
\frac{{\rm d} \Gamma^{(\epsilon)}_{\rm coll,2}}{d\s}
=
m_b^{-1}  \, \int_0^{1-\s} \, \frac{dx}{\bar x} \; 
 \int_0^1 d\s_{1s} \, \int_0^1 d\s_{2s} \, M_3 (\s/\bar x,\s_{1s},\s_{2s})
 \; f^{(\epsilon)}_\gamma (x,E_1) \; 
\left|\me{P_9}_{\rm tree} \right|^2_{\sot \rightarrow \s/\bar x} ,
\eea
where $\bar x=1-x$, and the non-linear change of variables $\sot \rightarrow
\s/\bar x$ also implied a distortion of the $x$-integration domain. The
addition of $\frac{{\rm d} \Gamma^{(\epsilon)}_{\rm coll,3}}{d\s} - \frac{{\rm
    d} \Gamma^{(\epsilon)}_{\rm coll,2}}{d\s}$ to the results of previous
subsections removes the remaining $\epsilon$-pole from the differential decay
width.

We are now free to convert back this observable to the usual one (in
which $\s$ is always the dilepton invariant mass) using mass
regularization. To this extent, we need the splitting function in this
scheme~\cite{Terazawa:1973tb}:
\be
\hskip -1cm
f^{(m)}_\gamma (x,E) = 4 \tilde \alpha_e \Big[
\frac{1+(1-x)^2}{x} 
\left(\ln\frac{E}{m_\ell}+\ln (2-2x) \right)
- \frac{1-x}{x}  - \frac{x}{2} \ln x - \frac{(2-x)^2}{2x} \ln(2-x)
\Big] \;.
\ee
The original differential decay width is then obtained by by adding
$\frac{{\rm d} \Gamma^{(m)}_{\rm coll,2}}{d\s} - \frac{{\rm d}
\Gamma^{(m)}_{\rm coll,3}}{d\s}$ where ${\rm d} \Gamma^{(m)}$ is
obtained from ${\rm d} \Gamma^{(\epsilon)}$ via $f_\gamma^{(\epsilon)}
\rightarrow f_\gamma^{(m)}$. Therefore, the total correction term is
given by the following double difference:
\bea
\frac{T_S}{2 m_b} & = &
\left(
\frac{{\rm d} \Gamma^{(m)}_{\rm coll,2}}{d\s} -
\frac{{\rm d} \Gamma^{(\epsilon)}_{\rm coll,2}}{d\s}
\right)
-
\left(
\frac{{\rm d} \Gamma^{(m)}_{\rm coll,3}}{d\s} -
\frac{{\rm d} \Gamma^{(\epsilon)}_{\rm coll,3}}{d\s}
\right) \; .
\eea
Note that only the $E$-independent difference $f^{(\epsilon)}_\gamma
(x,E) - f^{(m)}_\gamma (x,E)$ enters in the total correction term.
Hence, we can perform separately the ($x$, $\s_{12}$) and ($\s_{1s}$,
$\s_{2s}$) integrations. The tree level squared matrix element of
$P_9$ integrated over the phase space reads
\be
\sigma(\sot) \equiv 
\frac{2^{-7+6\epsilon} \pi^{-\frac{5}{2}+2\epsilon}(m_b^2)^{3-2\epsilon}
\tilde\mu^{4\epsilon}}{\Gamma(\frac{3}{2}-\epsilon)} d\sot \; \sot^{-\epsilon} \; 
(1-\sot)^{2-2\epsilon} \;
\frac{\Gamma(2-\epsilon)}{\Gamma(2-2\epsilon)} \; \frac{2\sot(1-\epsilon)+1}{3-2\epsilon}
\ee 
and the total correction term is finally expressed as
\bea
T_S & = & 2 \,  \left[ 
\int\limits_0^1 \!\! dx \;
\left[f^{(\epsilon)}_\gamma (x)- f^{(m)}_\gamma (x)\right]  \sigma (\s)
-
\int\limits_0^{1-\s} \!\! dx \;
\frac{f^{(\epsilon)}_\gamma (x)- f^{(m)}_\gamma (x)}{(1-x)} 
\sigma \! \left(\frac{\s}{1-x}\right) 
\right] 
\label{TS}
\eea
Both integrals in Eq.~(\ref{TS}) are infrared divergent for
$x\rightarrow 0$, but their sum is not.

The sum $T_V+T_R+T_S$ is now free of divergences and contains an explicit
collinear logarithm $\ln (m_b^2/m_\ell^2)$. The coefficient of this logarithm
vanishes when integrated over $\sot$. This means that if we had considered the
total branching ratio instead of the differential one, the sum of $T_V+T_R$
would have been already finite and the inclusion of $T_S$ would have become
unnecessary.  However, the coefficient of the collinear logarithm is large and
positive for low $\sot$ and large and negative for high $\sot$.  Furthermore,
this term renders by far the major contribution to the electromagnetic
  corrections. 
In  the sum $T_V+T_R+T_S$ the coefficient of
$Q_d^2$ is up to a color factor proportional to the QCD-function
$\omega_{99}^{(1)}(\s)$ from Eq.~(\ref{omega9one}), providing another check
for our result. Inserting $Q_d = -1/3$ and $Q_l = -1 $ finally yields
\be T_V+T_R+T_S =
\frac{\aem \, m_b^6 \, \left(1-\sot\right)^2 \left(1 +
2\,\sot\right)}{24\,\pi^3} \omega_{99}^{\rm (em)}(\sot) \;,
\label{p9result}
\ee
with
\bea
\omega_{ 99 }^{\rm (em)} (\s)& = &
\ln \left(\frac{m_b^2}{m_\ell^2}\right)\,\left[ - \frac{1 + 4\,\s - 8\,\s^2}{
 6\,\left( 1 - \s \right)\,\left( 1 + 2\,\s \right) } 
+ \ln (1 - \s) 
- \frac{\left( 1 - 6\,\s^2 + 4\,\s^3 \right) \,\ln \s}{
   {2\,\left( 1 - \s \right) }^2\,\left( 1 + 2\,\s \right) } \right] 
\nonumber \\ && \hskip -1cm
- \frac{1}{9} \,Li_2(\s) + \frac{4}{27} \pi^2 -
\frac{37 - 3\,\s - 6\,\s^2}{72\,\left( 1 - \s \right)\,\left( 1 + 2\,\s \right)}  - 
  \frac{\left( 41 + 76\,\s \right) \,\ln (1 - \s)}{36(1 + 2\,\s)} \nonumber\\
&& \hskip -1cm
+ \left( \frac{6 - 10\,\s - 17\,\s^2 + 14\,\s^3}{
18\left( 1 -\s \right)^2\,\left( 1 + 2\,\s \right) } + 
\frac{17\,\ln (1 - \s)}{18} \right) \,\ln \s - 
  \frac{\left( 1 - 6\,\s^2 + 4\,\s^3 \right) \, \ln^2 \s}{ 
2\left( 1 -\s \right)^2\,\left( 1 + 2\,\s \right) } \; .
\label{omegaem}
\eea
The contribution that we have calculated can be effectively taken into account
via the following substitution:
\bea
\left| C_9 (\mu_b) \me{P_9}_{\rm tree} \right|^2  
& \Longrightarrow &
\left|  C_9 (\mu_b)  \me{P_9}_{\rm tree} \right|^2 
\left[ 1 + 8 \;  \aem \; \omega_{99}^{\rm (em)} (\hat s) \right]  \; .
\eea

\subsection{Other log-enhanced corrections}\label{sec:otheremcorr}
The QED corrections to the matrix elements of $P_i$ with $i\neq 9$ 
 contribute to the branching ratio at 
  order  $O(\as^3 \kappa^3)$.  Consequently, following the outline in
  Section~\ref{sec:branchingratio}, we  include those contributions that are
enhanced by an explicit $\ln(m_b/m_\ell)$.  The relevant terms in the
amplitude are
\bea
{\cal A} &\propto& 
\Big[ (C_2 + C_F \; C_1) \; \as  \kappa \; f_2 (\s) + C_9 \Big] \me{P_9}_{\rm tree}
+ C_{10} \me{P_{10}}_{\rm tree} + C_7^{{\rm eff}} \me{P_7}_{\rm tree}
\eea
where the $f_{ 2 }(\s)$ is defined in Eq.~(\ref{4melem}). Here we 
  have  dropped the NNLO QCD corrections to the matrix elements as well as
 the  terms proportional to the  small  penguin coefficients 
  $C_{i(Q)}$. After  squaring and under the assumption that $C_1$ and $C_2$
are real, we obtain
\bea
\hskip -1cm 
|{\cal A}|^2 & \propto &  
\left[
|C_9|^2 
+ \as^2  \kappa^2 \; (C_2 + C_F \; C_1)^2 \; |f_2 (\s)|^2
+ 2 \; \as \kappa \; {\rm Re} [f_2(\s) (C_2 + C_F \; C_1) C_9^* ] 
\right] \; \left|\me{P_9}_{\rm tree}\right|^2 \nnb \\
& & 
+ 2 \; {\rm Re} \Big[ C_7^{{\rm eff}} C_9^* + \as \kappa \; C_7^{{\rm eff}} (C_2 + C_F \; C_1) f_2^* (\s)\Big]
\; \me{P_7}_{\rm tree}\me{P_9}_{\rm tree}^* \nnb\\
& &
+ |C_7^{{\rm eff}}|^2 \; \left|\me{P_7}_{\rm tree}\right|^2
+ |C_{10}|^2 \; \left|\me{P_{10}}_{\rm tree}\right|^2 \; .
\eea 
The fully differential decay width in the collinear limit now yields
\bea
{\rm d} \Gamma_{\rm coll}^{(m)} (\sot,\s_{1s},\s_{2s},x)
& = & 
m_b^{-1}  \, f^{(m)}_\gamma (x,E_1) \,  \left| {\cal A} \right|^2 \, dPS_3  \, dx \;.
\label{fullydifferentialwidthpi}
\eea
These corrections are induced by collinear photon emission and are
given by
$ \frac{{\rm d} \Gamma^{(m)}_{\rm coll,2}}{d\s} - 
  \frac{{\rm d} \Gamma^{(m)}_{\rm coll,3}}{d\s}$
where we retain only the $\ln(m_b/m_\ell)$ term in
$f_\gamma^{(m)}(x,E)$. The result reads
\bea
\frac{{\rm d} \Delta \Gamma}{d\s} &=&
\frac{G_F^2 m_b^5}{48 \pi^3} |V_{tb} V_{ts}|^2 (1-\s)^2 \as \kappa 
\Bigg\{ 8\, (1+ 2\s) \Bigg[ 
|C_9|^2 \;  \omega_{99}^{\rm (em)}(\s)  
+|C_{10}|^2 \; \omega_{1010}^{\rm (em)} (\s) 
\nnb\\
& &
+\as\kappa \; {\rm Re} \left[(C_2 + C_F C_1) C_9^* \;  \omega_{29}^{\rm (em)}(\s) \right]
+\as^2 \kappa^2 \; (C_2 + C_F C_1)^2 \;  \omega_{22}^{\rm (em)} (\s)
\Bigg]
\nnb \\
& & 
+96 \, \Bigg[ 
\as \kappa \; {\rm Re} \left[C_7^{{\rm eff}} C_9^*\right] \; \omega_{79}^{\rm (em)}(\s) 
+\as^2\kappa^2 \; {\rm Re} \left[(C_2 + C_F C_1) \, C_7^{{\rm eff} \, *} \;  \omega_{27}^{\rm (em)}(\s) \right]
\Bigg] 
\nnb\\
& &
+8\, (4+\frac{8}{\s})\as^2 \kappa^2 \; |C_{7}^{{\rm eff}}|^2 \; \omega_{77}^{\rm (em)} (\s) \Bigg\} \; \; ,
\label{deltagamma}
\eea
where $\omega_{99}^{\rm (em)}(\s)$ was already found in the previous section.
The other $\omega$-functions read:
\bea \label{om1010em}
\omega_{1010}^{\rm (em)}(\s) & = & 
\ln \left(\frac{m_b^2}{m_\ell^2}\right)\,\left[ - \frac{1 + 4\,\s - 8\,\s^2}{
 6\,\left( 1 - \s \right)\,\left( 1 + 2\,\s \right) } 
+ \ln (1 - \s) 
- \frac{\left( 1 - 6\,\s^2 + 4\,\s^3 \right) \,\ln \s}{
   {2\,\left( 1 - \s \right) }^2\,\left( 1 + 2\,\s \right) } \right] \;,\\
\omega_{77}^{\rm (em)}(\s) & = & 
\ln \left(\frac{m_b^2}{m_\ell^2}\right)\,\left[\frac{\s}{2\,{\left( 1 - \s \right) }\,\left( 2 + \s \right) } + 
  \ln (1 - \s) - 
  \frac{\s\,\left( -3 + 2\,\s^2 \right) }{2\,{\left( 1 - \s \right) }^2\,\left( 2 + \s \right) }\,\ln (\s)\right] \;,\\
\omega_{79}^{\rm (em)}(\s) & = & \ln \left(\frac{m_b^2}{m_\ell^2}\right)\,\left[-\frac{1}{2\,( 1- \s)} + \ln (1 - \s) + \frac{\left( 
-1 + 2\,\s - 2\,\s^2 \right) }{2\,{\left( 1 - \s \right) }^2}\,\ln (\s)\right]\;,\\
\omega_{29}^{\rm (em)}(\s) & = & 
\ln \left(\frac{m_b^2}{m_\ell^2}\right)\,\left[\frac{\Sigma_1(\s)+ i \,\Sigma_1^I(\s)}{8 (1-\s)^2 (1+2\s)}\right] + \frac{16}{9} \,
\omega_{1010}^{\rm (em)}(\s)\,\ln\!\left(\frac{\mu_b}{5\gev}\right)\;,\\
\omega_{22}^{\rm (em)}(\s) & = & 
\ln \left(\frac{m_b^2}{m_\ell^2}\right)\,\left[\frac{\Sigma_2(\s)}{8 (1-\s)^2 (1+2\s)} +  
\, \frac{\Sigma_1(\s)}{9 (1-\s)^2 (1+2\s)}\ln\!\left(\frac{\mu_b}{5\gev}\right)\right] \nnb\\ & & \nnb \\
    &&+ \, \frac{64}{81} \; \omega_{1010}^{\rm (em)}(\s)\, \ln^2\!\left(\frac{\mu_b}{5\gev}\right)
\;,\\ & & \nnb\\
\omega_{27}^{\rm (em)}(\s) & = & 
\ln \left(\frac{m_b^2}{m_\ell^2}\right)\,\left[\frac{\Sigma_3(\s)+ i \, \Sigma_3^I(\s)}{96 (1-\s)^2}\right] + \frac{8}{9} \,
\omega_{79}^{\rm (em)}(\s) \, \ln\!\left(\frac{\mu_b}{5\gev}\right)\;.
\label{om79em}
\eea
The  functions $\Sigma_i$ have been evaluated numerically in the low-$\s$-region (for
fixed values of $m_b$ and $m_c$). They are accurately reproduced by the following fits:

\bea
\Sigma_1(\s) &=&  23.787 - 120.948\, \s + 365.373\, \s^2 - 584.206\, \s^3 \;,\\
\Sigma_1^I(\s) &=&  1.653 + 6.009\, \s - 17.080\, \s^2 + 115.880\, \s^3 \;,\\
\Sigma_2(\s) &=&  11.488 - 36.987\, \s + 255.330\, \s^2 - 812.388 \, \s^3 + 1011.791 \, \s^4\;,\\
\Sigma_3(\s) &=&  109.311 - 846.039\, \s + 2890.115\, \s^2 - 4179.072\, \s^3 \;, \\
\Sigma_3^I(\s) &=&  4.606 + 17.650\, \s - 53.244\, \s^2 + 348.069\, \s^3 \;.
\eea  \\[-1cm]

\section{  Collinear logarithms and angular cuts  }\label{sec:logdiscussion}
The explicit logarithm of the lepton mass signals the presence of a
collinear singularity whose appearance in the differential branching
ratio is strictly related to the definition of the dilepton invariant
mass. As explained in Sec.~\ref{NDRtomass}, this logarithm
disappears if all photons emitted by the final state on-shell leptons
are included in the definition of $s$: $(p_{\ell_1}+p_{\ell_2})^2
\rightarrow (p_{\ell_1}+p_{\ell_2}+p_\gamma)^2$.

Let us consider a cone (of angular opening $\theta$) around an on-shell lepton
with momentum $p_\ell$. For all photons emitted in this cone we have:
$m_\ell^2\leq(p_\ell + p_\gamma)^2 \leq \Lambda^2 \simeq 2 E_\ell^2 (1-\cos
\theta)$, where $E_\ell$ is the energy of the lepton  (usually of order
  $m_b$). Consequently the effect of including such photons in the
  reconstruction of the lepton momentum can be roughly approximated by
  replacing $m_\ell$ by some scale of order $\Lambda$ in the collinear
  logarithm.
  
  Both Babar and Belle include sufficiently collinear photons in the
  definition of the lepton momentum. However, the imposed cones are so narrow
  that the effect for the muons is negligible, i.e. the separation of muons
  and collinear photons is practically perfect \cite{private}. Thus, our
  expressions containing $\ln(m_b^2/m_\mu^2)$ are directly applicable in this
  case. 

For electrons, the situation is more  complicated. In  both experiments,
the cone is defined in the laboratory frame and has polar and azimuthal angles
around 45$\,$mrad and 5$\,$mrad, respectively. Hence, $\Lambda$ is of the same
order as $m_\mu$,  which makes the QED corrections for the electrons similar to
those for the muons. Nothing more precise can be said without applying
dedicated Monte Carlo routines that would take into account the experimental
setups in detail. 

%

\section{Formulae for the branching ratio \label{sec:masterBR}}
In Section~\ref{sec:branchingratio}, we have expressed the branching ratio in
terms of the quantity $\Phi_{\ell\ell}(\s)/\Phi_u$. In the present Section, we
express this quantity in terms of the low-scale Wilson coefficients and
various functions of $\s$ that arise from the matrix elements. The main
formula reads
\bea \label{mainmaster}
\f{\Phi_{\ell\ell}(\s)}{\Phi_u} &=& \sum_{i\leq j} 
{\rm Re} \left[ C_i^{\rm eff} (\mu_b) \; C_j^{{\rm eff}*} (\mu_b)  
                \; H_{ij} (\mu_b,\s) \right] \;,
\eea
where $C_i^{\rm eff} (\mu_b) \neq C_i (\mu_b)$ only for $i=7,8$ (see Eqs.
(\ref{c7eff}) and (\ref{c8eff})). The functions $H_{ij}(\mu_b,\s)$ can be
expressed analytically in terms of the coefficients $M_i^A$ listed in
Table~\ref{tab:hiA} and of the following building blocks
\bea
{\rm S}_{99} & = & (1-\s)^2 (1+2\s) \left\{ 
                   1 + 8\; \as \left[ \omega_{99}^{(1)} (\s) + u^{(1)} \right]
                     + \kappa \, u^{(\rm em)} + 8\; \as \kappa \; \omega_{99}^{(\rm em)} (\s) 
\right. \nnb \\ &&   \left.
   + 16\; \as^2 \left[ \omega_{99}^{(2)} (\s) + u^{(2)} + 4 u^{(1)}
                       \omega_{99}^{(1)}(\s) \right] \right\} 
\nnb \\ &&
+\; 6 \; \frac{\lambda_2}{m_b^2} \; (1-6 \s^2+ 4\s^3) \;,\\
{\rm S}_{77} & = & (1-\s)^2 (4+\frac{8}{\s}) 
                   \left\{ 1+ 8\; \as \left[ \omega_{77}^{(1)}(\s) + u^{(1)} \right]
                   + \kappa \, u^{(\rm em)} + 8\; \as \kappa \; \omega_{77}^{(\rm em)} (\s) \right\}
\nnb \\ &&
+ \; 24 \; \frac{\lambda_2}{m_b^2} \; (2\s^2-3)\;,\\
{\rm S}_{79} & = & 12 (1-\s)^2 \left\{ 1+ 8\; \as \left[ \omega_{79}^{(1)} (\s) + u^{(1)} \right]
                   + \kappa \, u^{(\rm em)} + 8\; \as \kappa \; \omega_{79}^{(\rm em)} (\s) \right\}
\nnb \\ &&
+\; 24 \; \frac{\lambda_2}{m_b^2} \; (1-6 \s+4\s^2) \;,\\
{\rm S}_{1010} & = & S_{99} + 8\; \as \kappa (1-\s)^2 (1+2\s)
\left[ \omega_{1010}^{(\rm em)} (\hat s) - \omega_{99}^{(\rm em)} (\hat s) \right] \;.
\eea
The functions $\omega_{ij}^{(k)}$ are listed in
Appendix~\ref{sec:loopfunctions}.  The functions $\omega_{ij}^{\rm (em)}$ have
been given in Eqs.~(\ref{omegaem}) and (\ref{om1010em})--(\ref{om79em}). The
numbers $u^{(1)} = (4\pi^2-25)/12$ and 
 $u^{(2)} \simeq 27.1 + \beta_0 u^{(1)} \ln(\mu_b/m_b)$ 
  originate from the QCD corrections to $b \to X_u e \bar{\nu}$
  \cite{vanRitbergen:1999gs}, while the quantity $u^{(\rm em)} = \f{12}{23}
  \left(\eta^{-1}-1\right)$ stands for the logarithmically-enhanced QED
  correction to this decay \cite{Sirlin:1981ie}.  The $S_{AA}$ include
  non-perturbative ${\cal O}(1/m_b^2)$ corrections that one finds by taking
  the limit $m_c \to 0$ in Eq.~(18) of Ref.~\cite{Buchalla:1998mt}.

The explicit expressions for the functions $H_{ij}$ read
 \be 
H_{ij} = \left\{ \begin{array}{ll}
{}~\sum~~ |M_i^A|^2 \;S_{AA} + \;{\rm Re} (M_i^7 M_i^{9*}) \;S_{79} +\Delta H_{ii}\;,
& \mbox{when~} i=j \\[-2mm]
{\!\!\!\scriptscriptstyle A=7,9,10}\\[2mm]
{}~\sum~~ 2 M_i^A M_j^{A*} \; S_{AA} 
+ \; \left(M_i^7 M_j^{9*} + M_i^9 M_j^{7*} \right) \; S_{79} +\Delta H_{ij}\;,
& \mbox{when~} i\neq j \\[-2mm]
{\!\!\!\scriptscriptstyle A=7,9,10}
\end{array}\right.
\ee 
It is assumed that all the products in Eq.~(\ref{mainmaster}) are expanded in
$\as$, $\kappa$ and $\lambda_2$, and that higher orders are neglected (see
Section \ref{sec:branchingratio}).  The quantities
\be
\Delta H_{ij} = b_{ij} + c_{ij} + e_{ij} 
\ee
that need to be included only for $i=1,2$ stand for additional bremsstrahlung
($b_{ij}$), non-perturbative ${\cal O}(1/m_c^2)$ corrections ($c_{ij}$) and
additional $\ln (m_b^2/m_\ell^2)$-enhanced electromagnetic corrections ($e_{ij}$). 
Specifically, the non-vanishing $e_{ij}$ that we include read
\bea
e_{22} & = & 8\,(1-\s)^2\; (1+ 2\s)\;\as^3 \kappa^3\; \omega_{22}^{\rm (em)}(\s) \nnb\\
e_{27} & = & 96\,(1-\s)^2 \,\as^3 \kappa^3 \;\omega_{27}^{\rm (em)}(\s) \nnb\\
e_{29} & = & 8\,(1-\s)^2\, (1+ 2\s)\; \as^2 \kappa^2 \;\omega_{29}^{\rm (em)}(\s) \nnb\\
e_{11} & = &  \fm{16}{9}  \; e_{22} \nnb\\
e_{12} & = &  \fm{8}{3}   \; e_{22} \nnb\\
e_{1j} & = &  \fm{4}{3}  e_{2j}, \hskip 1.5cm {\rm for \; j=7,9}.
\eea

The ${\cal O}(1/m_c^2)$ non-perturbative contributions were calculated in
Ref.~\cite{Buchalla:1997ky}
\bea
c_{27} &=& -  \as^2 \kappa^2  \frac{8\lambda_2}{9 m_c^2} (1-\s)^2
                  \frac{1+6\s-\s^2}{\s}\,  {\rm Re}\,F(r)  \;,\nnb\\
c_{29} &=& - \as \kappa \frac{8\lambda_2}{9 m_c^2} (1-\s)^2 (2+\s)\,  {\rm Re}\,F(r)  \;,\nnb\\
 c_{22}  &=& - \as \kappa \frac{8\lambda_2}{9 m_c^2} (1-\s)^2 (2+\s)\,  
                   {\rm Re}\left( F(r)  M_2^{9*} \right), \nnb\\
c_{11} & = &  -\fm{2}{9}  \; c_{22} \nnb\\
c_{12} & = &  \fm{7}{6}  \; c_{22} \nnb\\
c_{1j} & = &  -\fm{1}{6}  \; c_{2j}, \hskip 1.5cm  {\rm for \; j=7,9} 
\eea 
 where $r = 1/y_c = m_{\ell^+ \ell^-}^2/(4 m_c^2)$. 
The function  $F(r)$  can be found in Appendix~\ref{sec:loopfunctions}. 

The finite bremsstrahlung contributions $b_{ij}$ appear at
NNLO  in Ref.~\cite{Asatryan:2002iy}, where the notation is very similar to the one
proposed here. We do not present these corrections here but do include them in
the numerical analysis.

\section{Acknowledgments}

We would like to thank Piotr Chankowski for bringing the problem of
electromagnetic logarithms $\ln(M_H^2/M_L^2)$ to our attention. We are
grateful to the authors of Ref.~\cite{Bobeth:2003at} for a careful reading of
the manuscript and useful remarks. We are deeply indebted to  Thomas~Gehrmann and
 Aude~Gehrmann-De Ridder for many discussions and help on the phase space
integration.  We thank Jeffrey Berryhill and Akimasa Ishikawa for
  information concerning the treatment of collinear photons at BaBar and
  Belle.  This work was supported in part by the Schwei\-zerischer
Nationalfonds.  M.M. acknowledges support from the Polish Committee for
Scientific Research under the grant 2~P03B~078~26 and by the European
Community's Human Potential Programme under the contract HPRN-CT-2002-00311,
EURIDICE. Research partly supported by the Department of Energy under Grant
DE-AC02-76CH03000. Fermilab is operated by Universities Research
Association Inc., under contract with the U.S. Department of Energy. 

\appendix
\section{Various functions} \label{sec:loopfunctions}
The loop functions that appear in the text are:
\bea
A(x) &=& \f{-3 x^3 + 2 x^2 }{ 2 (x-1)^4 } \ln x + \f{ 8 x^3 + 5 x^2 - 7 x }{ 12 (x-1)^3 },\\[2mm]
Y(x) &=& \f{3 x^2 }{ 8 (x-1)^2} \ln x + \f{ x^2 - 4 x }{ 8 (x-1)},\\[2mm]
W(x) &=& \f{ -32 x^4 +  38 x^3 +  15 x^2 -  18 x }{ 18 (x-1)^4 } \ln x +
      \f{ -18 x^4 + 163 x^3 - 259 x^2 + 108 x }{ 36 (x-1)^3 },\\[2mm]
S(x) &=& \f{3 x^3}{ 2 (x-1)^3 } \ln x + \f{ x^3 - 11 x^2 + 4 x }{ 4 (x-1)^2 },\\[2mm]
X(x) &=& \f{3 x^2 - 6 x }{ 8 (x-1)^2 } \ln x + \f{ x^2 + 2 x }{ 8 (x-1) } ,\\[2mm]
E(x) &=& \f{x (18 -11
x - x^2)}{12 (1-x)^3} + \f{x^2 (15 - 16 x + 4 x^2)}{6 (1-x)^4} \ln
x-\f{2}{3} \ln x \; .
\eea

The following function appears in the matrix elements of the 4-quark operators:
\bea
g(y) & = &
\f{20}{27} + \f{4}{9} y - \f{2}{9}(2+y) \sqrt{|1-y|} \left\{ \begin{array}{ll}
\ln \left|\f{1+\sqrt{1-y}}{1-\sqrt{1-y}}\right| - i \pi, & {\rm when}~ y < 1,\\[2mm]
2 \arctan \f{1}{\sqrt{y-1}},                     & {\rm when}~ y \ge 1 \;,
\end{array} \right.
\eea

The $\omega_{ij}^{(n)}$ functions that include the sum of infrared divergent
virtual and real contributions to the matrix elements of $P_7$, $P_9$
and $P_{10}$ are:
\bea
\omega_{99}^{(1)} (\hat s) & = &
-\f{4}{3} Li_2(\s) -\f{2}{3} \ln(1-\s) \ln\s -\f{2}{9} \pi^2 
-\f{5+4\s}{3(1+2\s)} \ln(1-\s) \nonumber \\ 
&&-\f{2\s(1+\s)(1-2\s)}{3(1-\s)^2(1+2\s)}\ln \s +\f{5+9\s-6\s^2}{6(1-\s)(1+2\s)}\;,\label{omega9one}\\
\omega_{99}^{(2)}(\hat s) & = & \frac{-19.2 + 
6.1 \s + (37.9 + 17.2 \ln\s)\s^2 - 18.7 \s^3}{(1-\s)^2(1+2\s)} \;, \label{om299}\\
\omega_{1010}^{(1)} (\hat s) & = & \omega_{99}^{(1)} (\hat s)\;, \\
\omega_{77}^{(1)} (\hat s) & = & 
- \frac{4}{3}Li_2 (\s)- \frac{2}{3} \ln (1 - \s)\,\ln \s  -\frac{2}{9}\pi^2 
- \frac{\left( 8 + \s \right) }{3\,\left( 2 + \s \right) } \ln (1 - \s) 
\nonumber\\
& &
 - \frac{2\,\s\,\left( 2 - 2\,\s - \s^2 \right) 
}{3\,{\left( 1 - \s \right) }^2\,\left( 2 + \s \right) } \ln \s
- \frac{16 - 11\,\s - 17\,\s^2}{18\,\left( 1 - \s \right) \,
\left( 2 + \s \right) } - \frac{8}{3} \ln (\frac{\mu_b}{m_b}) 
\;, \\
\omega_{79}^{(1)}  (\hat s)& = &
- \frac{4}{3}Li_2 (\s)- \frac{2}{3} \ln (1 - \s)\,\ln \s  -\frac{2}{9}\pi^2 
- \frac{\left( 2 +7 \s \right) }{9 \s  } \ln (1 - \s) \nonumber\\
& &
 - \frac{2\,\s\,\left(3 - 2\s \right) 
}{9\,{\left( 1 - \s \right) }^2 } \ln \s 
+ \frac{5-9\s }{18\,\left( 1 - \s \right) } - \frac{4}{3} \ln (\frac{\mu_b}{m_b}) \;.
\eea
 The function  $\omega_{99}^{(1)}(\hat s)$  has been extracted
\cite{Misiak:1992bc,Buras:1994dj} from the ${\cal O}(\alpha_s)$ corrections
\cite{Jezabek:1988ja} to the semileptonic decay.  The functions
 $\omega_{77}^{(1)}(\hat s)$  and $\omega_{79}^{(1)}(\hat s)$ have been calculated
in Ref.~\cite{Asatryan:2001zw}. Note that  $\omega_{77}^{(1)}(\hat s)$ 
in the $\s\to 0$ limit reproduces the ${\cal O}(\alpha_s)$ correction
\cite{Ali:1990tj} to the matrix element of $P_7$ in the $b \to X_s \gamma$
decay.  The function $\omega_{ 99 }^{(2)}(\hat s)$ was extracted
  \cite{Bobeth:2003at} from the ${\cal O}(\alpha_s^2)$ corrections
  \cite{Chetyrkin:1999ju,Czarnecki:2001cz} to the spectrum of the $b \to X_u e
  \bar\nu$ decay. The approximate formula in Eq.~(\ref{om299}) is 
valid in the range $0<\s<0.4$.

The function $F(r)$ that arises in the ${\cal O}(1/m_c^2)$
non-perturbative corrections reads \cite{Buchalla:1997ky}
\be
F(r) = \f{3}{2r} \left\{ 
\begin{array}{ll}
\f{1}{\sqrt{r(1-r)}} \arctan \sqrt{\f{r}{1-r}}~-1, & {\rm ~when~} 0 < r < 1,\\
\f{1}{2\sqrt{r(r-1)}} \left( \ln \f{1-\sqrt{1-1/r}}{1+\sqrt{1-1/r}} +i\pi\right) -1 ,
& {\rm ~when~} r > 1.
\end{array}
\right.
\ee   

\section{$\hat V$ and $a_i$}\label{app:Va}

The numerical diagonalization of the matrix $\hat W^{(10)}$ yields:
\bea
a_i & = & \Big[-1.04348,-0.899395,-0.521739,-0.521739,-0.422989,0.408619, \nonumber\\
 & &\hskip 4cm  0.26087,0.26087,0.26087,0.145649,0.130435,0,0 \Big]
\eea
and
\bea
\hat V & = & 
\left( \matrix{
\se 0 & \se 0 & \se 0.942522 & \se 0.0253179 & \se 0 & \se 0  \cr
\se 0 & \se 0 & \se -0.314174 & \se -0.0084393 & \se 0 & \se 0    \cr
\se -0.0109144&\se -0.160583&\se 0.0349082&\se -0.0961354&\se 0.917797&\se -0.922049  \cr
\se -0.0654862 & \se -0.984073 & \se -0.104725 & \se 0.288406 & \se -0.266582 & \se 0.331368 \cr
\se 0.000682148 & \se 0.00725171 & \se -0.00872705 & \se 0.0240338 & \se -0.153681 & \se 0.130848 \cr
\se 0.00409289 & \se 0.0759058 & \se 0.0261812 & \se -0.0721015 & \se 0.250927 & \se 0.151325  \cr 
\se 0 & \se 0 & \se 0 & \se 0 & \se 0 & \se 0  \cr
\se 0 & \se 0 & \se 0 & \se 0 & \se 0 & \se 0   \cr
\se 0.163715 & \se 0 & \se 0 & \se 0.291219 & \se 0 & \se 0   \cr
\se 0.982293 & \se 0 & \se 0 & \se -0.873658 & \se 0 & \se 0   \cr
\se -0.0102322 & \se 0 & \se 0 & \se -0.0728048 & \se 0 & \se 0    \cr
\se -0.0613933 & \se 0 & \se 0 & \se 0.218414 & \se 0 & \se 0    \cr
\se 0 & \se 0 & \se 0 & \se 0 & \se 0 & \se 0   \cr
} \right.\nonumber \\
& &
\hskip 2cm 
\left. \matrix{
\se 0.00100213 & \se -0.83105 & \se 0.00542193 & \se 0 & \se 0 & \se 0 & \se 0 \cr 
 \se 0.00066809 &\se -0.554033 & \se 0.00361462 & \se 0 & \se 0 & \se 0 & \se 0 \cr 
\se -0.0255649 &\se -0.0263825 & \se 0.0632231 & \se 0.726443 & \se 0.0531116 & \se 0 & \se 0 \cr 
\se -0.0383473&\se -0.0395738 & \se 0.0948347 & \se -0.684418 & \se -0.0398337 & \se 0 & \se 0 \cr 
\se 0.00639122&\se 0.00659563 & \se -0.0158058 & \se -0.0368909 & \se -0.00331947 & \se 0 & \se 0 \cr 
 \se 0.00958682&\se 0.00989345 & \se -0.0237087 & \se 0.0499047 & \se 0.00248961 & \se 0 & \se 0 \cr 
 \se 0 &\se 0 & \se 0 & \se 0 & \se 0 & \se 1. & \se 0 \cr 
 \se 0 &\se 0 & \se 0 & \se 0 & \se 0 & \se 0 & \se 1. \cr
 \se -0.53753&\se 0 & \se 0 & \se 0 & \se -0.796674 & \se 0 & \se 0 \cr 
 \se -0.806295 &\se 0 & \se 0 & \se 0 & \se 0.597505 & \se 0 & \se 0 \cr 
 \se 0.134383&\se 0 & \se 0 & \se 0 & \se 0.0497921 & \se 0 & \se 0 \cr 
 \se 0.201574&\se 0 & \se 0 & \se 0 & \se -0.0373441 & \se 0 & \se 0 \cr 
 \se 0&\se 0 & \se 0.993053 & \se 0 & \se 0 & \se 0 & \se  0 \cr  } \right)
\eea

\section{Details of the bremsstrahlung calculation}
\label{app:details}

This last appendix is devoted to some technical details of the
bremsstrahlung calculation. We will integrate the sample kernel  $K_R =
\s_{1q}^{-1} \s_{2q}^{-1}$  over the four particle phase
space, show how the Gram determinant factorizes, and explain how to
extract all terms up to and including order $\epsilon^0$
analytically. Omitting bothersome prefactors and, in addition,
removing all hats from the invariants $\s_{ij}$ we consider the
expression
\bea\label{appdefiniton}
A &:= & ds_{12} \int\limits_{0}^{1}\, ds_{1s} \, ds_{1q} \, ds_{2q} \, ds_{2s} \, ds_{sq}   
\delta(1-s_{12}-s_{1s}-s_{1q}-s_{2s}-s_{2q}-s_{sq}) \nnb \\
&& \times \left(-\Delta_4\right)^{-\frac{1}{2}-\epsilon}    \Theta(-\Delta_4)    s_{1q}^{-1} s_{2q}^{-1}  \; ,
\eea
where the Gram determinant is given by
\be
\Delta_4 = (s_{12} s_{sq})^2 + (s_{1s} s_{2q})^2 +
(s_{1q}s_{2s})^2 - 2 \, (s_{12} s_{1s} s_{2q} s_{sq} + s_{1s}
s_{1q}
s_{2s} s_{2q} + s_{12} s_{1q} s_{2s} s_{sq}).
\ee
The first integration is over the $\delta$-function.  It is  
done by means of the variable $s_{sq}$, yielding
\bea\label{appdeltafunc}
A &= & ds_{12} \hspace*{-15pt} \int\limits_{0}^{\hspace*{15pt}1-s_{12}} \hspace*{-15pt} ds_{1s}
\hspace*{-35pt}\int\limits_{0}^{\hspace*{35pt}1-s_{12}-s_{1s}} \hspace*{-35pt}ds_{1q}    s_{1q}^{-1}
\hspace*{-50pt}\int\limits_{0}^{\hspace*{55pt}1-s_{12}-s_{1s}-s_{1q}} \hspace*{-55pt}ds_{2q}    s_{2q}^{-1}
\hspace*{-58pt}\int\limits_{0}^{\hspace*{75pt}1-s_{12}-s_{1s}-s_{1q}-s_{2q}} \hspace*{-65pt}ds_{2s} \nnb \\
&& \hspace*{140pt}\times \left(-\Delta_4\right)^{-\frac{1}{2}-\epsilon}    \Theta(-\Delta_4) \; _{\big|
s_{sq}=1-s_{12}-s_{1s}-s_{1q}-s_{2q}-s_{2s}} \; .
\eea
Substituting $s_{sq}=1-s_{12}-s_{1s}-s_{1q}-s_{2q}-s_{2s}$ in the Gram
determinant yields an object that can be transformed into a quadratic
polynomial in either of the variables $s_{1s}$, $s_{1q}$, $s_{2s}$ or
$s_{2q}$, \textit{i.e.} in either variable that does not accompany $s_{sq}$ in
the quadratic piece of the Gram  determinant. We choose this quadratic
polynomial to be in $s_{2s}$:
\be\label{appquadratpoly}
- \Delta_4 = - (s_{12}+s_{1q})^2    \left[s_{2s}^2 + 2 \, B \, s_{2s} + C\right] = (s_{12}+s_{1q})^2    (s_{2s}^{+}-s_{2s})   
(s_{2s}-s_{2s}^{-}) \; ,
\ee
where $s_{2s}^{\pm}$ are the roots of the quadratic polynomial:
\be\label{approots}
s_{2s}^{\pm} = - B \pm \sqrt{B^2-C} \equiv -B\pm \sqrt{\Xi}.
\ee
The $\Theta$-function now requires these roots to be
real\footnote{otherwise $-\Delta_4$ is negative for all $s_{2s}$}
which is equivalent to the condition $\Xi \ge 0$.  From the latter
inequality we conclude
\be\label{appzuplim}
s_{2q} \le z    \left(1-s_{12}-s_{1s}-s_{1q}\right) \qquad \mbox{with} \qquad z = \frac{s_{12}+s_{1q}}{s_{12}+s_{1q}+s_{1s}} \le 1
\ee
The above roots now fulfill the inequality $0 \le s_{2s}^{-} \le
s_{2s}^{+} \le 1-s_{12}-s_{1s}-s_{1q}-s_{2q}$ which leads to the
following new limits of integration:
\bea\label{appnewlimits}
A &= & ds_{12} \hspace*{-15pt} \int\limits_{0}^{\hspace*{15pt}1-s_{12}} \hspace*{-15pt} ds_{1s}
\hspace*{-35pt}\int\limits_{0}^{\hspace*{35pt}1-s_{12}-s_{1s}} \hspace*{-35pt}ds_{1q}    s_{1q}^{-1}   
(s_{12}+s_{1q})^{-1-2\epsilon} \nnb \\
&& \times \hspace*{-55pt} \int\limits_{0}^{\hspace*{55pt}z (1-s_{12}-s_{1s}-s_{1q})} \hspace*{-55pt}ds_{2q}    s_{2q}^{-1}
\hspace*{17pt}\int\limits_{s_{2s}^{-}}^{s_{2s}^{+}} ds_{2s} \; (s_{2s}^{+}-s_{2s})^{-\frac{1}{2}-\epsilon}\,
(s_{2s}-s_{2s}^{-})^{-\frac{1}{2}-\epsilon} \; .
\eea
Substituting $s_{2s} = (s_{2s}^{+}-s_{2s}^{-})    \chi +
s_{2s}^{-}$, the subsequent $\chi$-integration can be done trivially
in terms of $\Gamma$-functions.

As a general strategy for the choice of the order of integration, we suggest
the following. The variable of the first integration ($\delta$-function) must
not be contained in the term of $K_R$ that one considers. If possible, this
term should also be free of the variable that one uses to factorize the Gram
 determinant  ($s_{2s}$~in our case). If the latter is not possible as
for instance in $K_R = s_{1s} s_{2s} s_{1q}^{-1} s_{2q}^{-1}$, 
one should at least factorize the Gram  determinant  in a
variable that does not appear in the denominator of $K_R$. This procedure
ensures that the first two integrations can be done in terms of
$\Gamma$-functions, and it avoids hypergeometric functions to emerge at this
stage of the calculation.

The choice of the subsequent order of integration is governed by the
aim to extract all divergences as early as possible, this being the
reason why we solved the condition $\Xi \ge 0$ for $s_{2q}$. We now
substitute $s_{2q} = z (1-s_{12}-s_{1s}-s_{1q}) \, t$ and perform the
$t$-integration, yielding again only $\Gamma$-functions. After
simplification, we obtain
\be\label{appaftertint}
\hskip -1cm 
A =  -\frac{\pi}{\epsilon} \, ds_{12} \, s_{12}^{-\epsilon}\hspace*{-15pt} \int\limits_{0}^{\hspace*{15pt}1-s_{12}}
\hspace*{-15pt} ds_{1s}  \, s_{1s}^{-\epsilon} \hspace*{-35pt}\int\limits_{0}^{\hspace*{35pt}1-s_{12}-s_{1s}} \hspace*{-35pt}ds_{1q}  
s_{1q}^{-1-\epsilon}   (s_{12}+s_{1q})^{-1}   (s_{12}+s_{1q}+s_{1s})^{\epsilon}   (1-s_{12}-s_{1s}-s_{1q})^{-2\epsilon}\; .
\ee
We now proceed as follows
\begin{itemize}
\item Perform an expansion into partial fractions via $\displaystyle
\frac{1}{s_{1q}(s_{12}+s_{1q})} = \frac{1}{s_{12}\,s_{1q}} -
\frac{1}{s_{12}(s_{12}+s_{1q})}$
\item Substitute $s_{1q}=(1-s_{12}-s_{1s})   (1-u) \, .$ The
$u$-integration can be carried out in the first term of the above
expansion.
\item Substitute $s_{1s}=(1-s_{12})   (1-v) \, ,$ again the
$v$-integration can be done in the first term.
\end{itemize}

 One now obtains the following expression:
\bea\label{appfinally}
A 
&=&  
\hskip -2.4pt
-\frac{\pi}{\epsilon} \, ds_{12} \, s_{12}^{-1-\epsilon}  (1-s_{12})^{1-4\epsilon}
\Bigg\{\frac{\Gamma(1-2\epsilon)\Gamma(-\epsilon)\Gamma(1-\epsilon)}{\Gamma(2-4\epsilon)} \,
_2F_1(-\epsilon,1-2\epsilon;2-4\epsilon;1-s_{12}) \nnb \\
&&
-(1-s_{12}) \int\limits_0^1 \! du \int\limits_0^1 \! dv \, \frac{u^{-2\epsilon} (1-u)^{-\epsilon} \, v^{1-3\epsilon}
(1-v)^{-\epsilon}}{s_{12}+(1-s_{12}) \, v \, (1-u)} \left[1-(1-s_{12}) \, u \, v\right]^{\epsilon}\Bigg\}\, .
\eea
We now carry out a two-dimensional variable transformation from $(u,v)$ to $(y,w)$ via
\be
y \, = \, 1 - u \, v \qquad {\rm and} \qquad y \, w = v \, (1-u) \; .
\ee
The $w$-integration can now be performed, resulting in another hypergeometric
function. After using the Kummer relation, also the $y$-integration can be
done. The final result for $A$ reads
\bea\label{appfinally2}
A 
&=&  
\hskip -2.4pt
-\frac{\pi}{\epsilon} \, ds_{12} \, s_{12}^{-1-\epsilon}  (1-s_{12})^{1-4\epsilon}
\Bigg\{\frac{\Gamma(1-2\epsilon)\Gamma(-\epsilon)\Gamma(1-\epsilon)}{\Gamma(2-4\epsilon)} \,
_2F_1(-\epsilon,1-2\epsilon;2-4\epsilon;1-s_{12}) \nnb \\
&&
-(1-s_{12}) \, \frac{\Gamma^2(1-\epsilon)\Gamma(1-2\epsilon)}{\Gamma(3-4\epsilon)} \,
_2F_1(1-\epsilon,2-2\epsilon;3-4\epsilon;1-s_{12})\Bigg\}\, .
\eea
All the divergences have now been extracted in terms of poles and
$\Gamma$-functions. The remaining task is now to expand the hypergeometric
functions in $\epsilon$. This can can be done by means of the {\tt
  Mathematica} package {\tt HypExp} \cite{Huber:2005yg}. 

\setlength {\baselineskip}{0.2in}
 
\end{document}